\documentclass[aps,prl,superscriptaddress, twocolumn, showpacs]{revtex4-1}
\usepackage[utf8]{inputenc}
\usepackage{bm}
\usepackage{mathtools}
\usepackage{bbold}
\usepackage{amsmath}
\usepackage{amssymb}
\usepackage{color}
\usepackage[dvipsnames]{xcolor}
\usepackage{multirow}
\usepackage{comment}
\usepackage{cancel}
\usepackage{float}
\usepackage{color}
\usepackage[dvipsnames]{xcolor}
\usepackage{epstopdf}
\usepackage{amsmath}
\usepackage{braket}
\usepackage{amsthm}
\usepackage{amssymb}
\usepackage{amsfonts}
\usepackage{graphicx,subfigure}
\usepackage{txfonts}
\usepackage{ulem}
\usepackage{mathtools}
\usepackage{bm}
\usepackage{bbm}
\usepackage{hyperref}
\usepackage{natbib}
\DeclareUnicodeCharacter{2009}{\,}

\begin{document}

\newcommand{\INO}{Istituto Nazionale di Ottica, Consiglio Nazionale delle Ricerche (INO-CNR), Largo Enrico Fermi 6, 50125 Firenze, Italy}
\newcommand{\LENS}{European Laboratory for Nonlinear Spectroscopy (LENS), Via N. Carrara 1, 50019 Sesto Fiorentino, Italy}
\newcommand{\UNIFI}{Physics Department, University of Firenze, Italy}
\newcommand{\QSTAR}{QSTAR, Largo Enrico Fermi 2, 50125 Firenze, Italy}
\newcommand{\be}{\begin{equation}}
\newcommand{\ee}{\end{equation}}

\title{Programmable entanglement between atomic clouds in a cavity for quantum-enhanced differential measurements}
\title{Programmable Cavity Squeezing for Distributed Sensing in a Tweezer Array}

\author{Youssef Trifa}
\email{youssef.trifa@ino.cnr.it}
\affiliation{\INO}
\affiliation{\LENS}
\author{Marco Fattori}
\affiliation{\INO}
\affiliation{\LENS}
\affiliation{\UNIFI}
\author{Luca Pezz\`e}
\email{luca.pezze@ino.cnr.it}
\affiliation{\INO}
\affiliation{\LENS}
\date{\today}
	
\begin{abstract}
Field sensing with state-of-the-art atom interferometers is restricted to the use of uncorrelated devices operating in parallel. 
We can overcome this limitation by using distributed sensing protocols where quantum correlations among spatially-separated devices are engineered in the spatial mode carrying the signal.  
We show that a tweezer array in a cavity offers an ideal testbed to engineer quantum states for distributed sensing, with the possibility to generate 
entanglement both within and between the clouds. 
The competition between local and intercloud cavity-mediated exchange allows the sign and spatial pattern of the intercloud couplings to select the squeezed mode.
For two ensembles, positive coupling produces uniform collective squeezing, whereas negative coupling generates strong staggered, nonlocal squeezing. 
A semiclassical analysis reveals a counter-twisting-like phase-space flow, qualitatively distinct from standard one-axis twisting.
The analysis and results can be further generalized to a larger number of ensembles.
We apply the scheme to differential Ramsey interferometry with common phase noise spanning the full $2\pi$ range, the resulting staggered states reduce the phase uncertainty below the standard quantum limit, with an ellipse estimator approaching the Cramér–Rao bound. 
These results establish programmable cavity interactions as a scalable route to entanglement tailored to distributed signals.
\end{abstract}

\maketitle

\vspace{1cm}

{\it Introduction.---}
Atom interferometers~\citep{CroninRMP2009,Poli2013,LudlowRMP2015} are promising platforms for deployable quantum technologies~\citep{BongsNatRev2019}.
By the coherent manipulation of atomic ensembles, they convert weak perturbations into measurable phase shifts with exceptional sensitivity and long-term stability.
State-of-the-art devices are approaching the standard quantum limit (SQL) set by uncorrelated atoms~\citep{RosiNature2014, Janvier2022}. 
Surpassing this limit at fixed atom number requires quantum correlations~\cite{PezzeRMP2018}, most notably spin squeezing~\citep{ Salvi2018, Szigeti2020, CorgierPRL2021}.
Entanglement-enhanced sensitivity has been demonstrated across atomic clocks~\citep{Colombo2022, RobinsonNatPhys2024, Yang2025}, magnetometers~\citep{OckeloenPRL2013, Muessel2014}, and atom interferometers~\citep{Greve2022, Cassens2025}.
Yet, most atomic sensing architectures still treat spatially separated devices as independent probes, each estimating a local scalar parameter.
This paradigm is reaching its conceptual limitations~\citep{KomarNATPHYS2014, malia_nat_2022, Nichol2022, LiScience2026, Dietze2026}: several emerging applications require the reconstruction of signals distributed across space and time~\cite{YeZollerPRL2024, PezzeArXiv2025}.
Examples include gradient magnetometry~\cite{Apellaniz2018, malia_nat_2022}, gravity gradiometry and cartography~\cite{Stray2022}, multiaxis inertial navigation~\cite{Dickerson2013, Templier2022,  Stolzenberg2025}, and differential clock comparisons~\cite{Nichol2022, Zheng2022, EcknerNATURE2023, Dietze2026}.
These tasks require precise estimation of spatial phase combinations while possibly rejecting common-mode fluctuations~\citep{BaynhamNature2026, KitsonPRR2026}.
Spin squeezing of local atomic ensembles alone is therefore not generally optimal~\cite{gessner_multiparameter_2020}: quantum correlations must instead be engineered directly in the spatial mode carrying the signal~\cite{Pezz2025, Bate2025}.

Arrays of optical tweezers embedded in optical cavities offer precisely
this capability, with recent experiments demonstrating site-resolved
extended arrays collectively coupled to a common high-cooperativity
cavity mode~\cite{Picot2026}.
Tweezers provide local control of spatially separated atomic ensembles~\cite{CooperPRX2018, YoungNat2020, KaufmanNatPhys2021, ManteschNat2025}, while cavity photons mediate programmable long-range interactions between them~\citep{Periwal2021, GlicensteinPRA2021, Cooper2024, SeubertPRX2025}.
Crucially, these interactions can be switched off during sensing, allowing nonlinear state preparation and readout~\cite{Davis2016, Frowis2016, Macri2016, Hosten2016Nat}, while preserving linear parameter encoding.
Recent work has proposed coherent and dissipative cavity-mediated preparation of entangled states for differential sensing, including Lieb--Mattis dynamics in a decoherence-free subspace~\citep{Kaubruegger2026} or through chiral dissipative couplings in one-dimensional arrays~\citep{Chu2026}.
Here we pursue a complementary Hamiltonian-engineering strategy that extends cavity squeezing beyond a single uniform collective mode~\citep{LerouxPRL2010,HostenScience2016b,PedrozoNature2020}.
Starting from uncorrelated ensembles, programmable cavity dynamics generates nonlocal squeezing directly in a selected spatial mode, thereby structuring the entanglement to the specific distributed sensing task.

\begin{figure}[h!]
\centering
\includegraphics[width=0.4\textwidth]{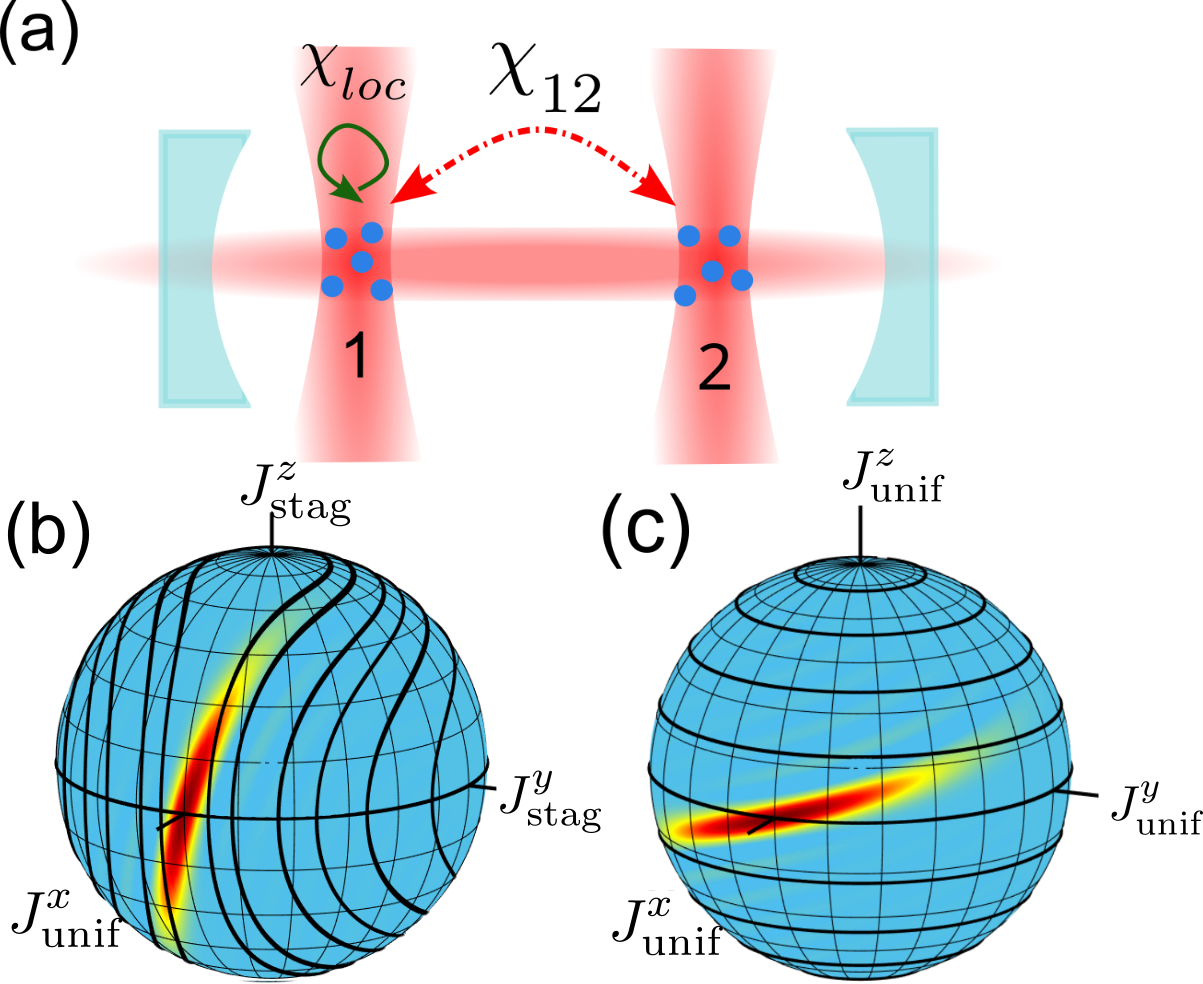}
\caption{
Geometric configuration of the system studied: two atomic clouds are trapped within an optical cavity. We approximate each atom as a two-level system, and construct spin operators for each cloud. Local and long-range interactions lead to the Hamiltonian described in Eq. \ref{hamM} for $M=2$. (b-c) Wigner distribution of the evolved state after a quench protocol from a coherent spin state along $x$ with (b) $\chi_{12} = -2 \chi_{loc}$ and (c) $\chi_{12} = 2 \chi_{loc}$, with $N_{1} = N_{2} = 20$ atoms in each cloud. Black lines indicates the semiclassical trajectories, see main text for more details. 
Note that in (b) the axes correspond to the staggered collective spin -one of its components being uniform to ensure standard spin commutation relations-, while in (c) they correspond to the uniform one.}
\label{fig1}
\end{figure}

We consider an array of $M$ atomic ensembles trapped in optical tweezers inside a common cavity, where local one-axis twists (OAT)~\cite{kitagawa_squeezed_1993} competes with cavity-mediated exchange interactions.
The sign and the spatial pattern of the exchange couplings program the collective mode in which quantum fluctuations are squeezed.
For two ensembles, illustrated in Fig.~\ref{fig1}, the interplay of local and non-local interaction produces two distinct regimes: positive coupling reduces the dynamics to conventional OAT of the uniform collective spin, whereas negative coupling drives a counter-twisting-like evolution in the staggered sector.
The latter directly generates strong nonlocal squeezing in the quadrature relevant to differential sensing.
We generalize this mode-selective mechanism to larger arrays, providing a programmable route to spatially structured entanglement for differential sensors and atomic-clock networks.
This coherent approach offers an alternative to protocols based on spatially split entangled Bose--Einstein condensates~\citep{LiScience2026}, quantum-non-demolition~\citep{malia_nat_2022}, and spin-squeezing transfer or swapping~\citep{Corgier2023, Malitesta2023, Mamaev2025}.
As an application, we study differential Ramsey interferometry under common phase noise uniformly distributed over the full $2\pi$ range.
The cavity-generated staggered states reject these common fluctuations while retaining sensitivity to the differential phase, reducing the estimation uncertainty to nearly half of the SQL for an array of $M=20$ ensembles.

{\it The model.---}
The $j$th ensemble in the cavity ($j=1, ..., M$) is composed by $N_j$ two-level atoms and it is modeled as a collective spin $S_{j} = (S_j^x, S_j^y, S_j^z)$ of length $N_{j}/2$.
The local spin operator $S^{\alpha}_{j}$ are constructed from the sum of atomic spin operators in each cloud, each atom being an effective spin-1/2. 
Taking into account light-induced interactions from the cavity mode~\cite{Kaubruegger2026, Periwal2021, Chu2026} (see Supplemental Material (SM) A), it is possible to model the system by an XY-exchange Hamiltonian supplemented by local one-axis-twisting nonlinearities,
\begin{equation}
    \label{hamM}
    \hat{H} = - \sum_{i<j} \chi_{ij} \big( S^{x}_{i} S^{x}_{j} + S^{y}_{i} S^{y}_{j}\big) + \chi^{\rm loc}_j  \sum_{i=1}^{M}  \big( S^{z}_{j} \big)^{2}. 
\end{equation}
This Hamiltonian has two tunable parameters: $\chi_{loc}$, which governs local interactions of atoms within the same cloud, and $\chi_{ij}$, which describes interaction between spatially-separated ensembles $i$ and $j$. 
Both parameters come from the interaction between the atoms and the cavity mode, and can be tuned independently by adjusting the laser power and the detuning with respect to transitions with a third atomic level. 
In particular, $\chi_{ij}$ does not need to be uniform across all clouds and can be tuned to engineer different interaction patterns~\cite{Periwal2021}.

In the following, we will explore the competition between these two terms on the entanglement dynamics between the atomic clouds.
We will consider independent initial ensembles, each prepared in a coherent spin state (CSS).
In particular, we compute the time evolution of the squeezing matrix~\cite{gessner_multiparameter_2020}, which is a generalization of the Wineland squeezing parameter~\cite{winelandPRA1994} in the context of multiparameter estimation (see Appendix and SM B), from which we extract the collective uniform and staggered squeezing parameters, $\xi^{2}_{\rm unif}$ and $\xi^{2}_{\rm stag}$ respectively.
The uniform squeezing corresponds to squeezing in the multi-ensemble spin $J_{\rm unif} = (J^{x}_{\rm unif}, ~J^{y}_{\rm unif}, ~J^{z}_{\rm unif})$, with $J^{\mu}_{\rm unif} = \sum_{j} S^{\mu}_{j}$ being the uniform  collective spin along $\mu$.
In contrast, the staggered corresponds to squeezing of $J_{\rm stag} = (J^{x}_{\rm unif}, ~J^{y}_{\rm stag}, ~J^{z}_{\rm stag})$, with  $J^{\mu}_{\rm stag} = \sum_{j} (-1)^{j+1} S^{\mu}_{j}$.
For simplicity, in the following, we restrict to the case where atom number is the same in each ensemble, $N_j=N$, and $\vert \chi_{ij}\vert= \chi$. We will use $S$ to refer to the local spins and $J$ for the collective one.
The inequality $\xi^{2}_{\rm unif}<1$ certifies usefulness of the state to estimate sum of phase shifts $\theta_1 + \theta_{2}+ \dots$ when applying a rotation $e^{-i S_z \theta_j}$ to each ensemble, while $\xi^{2}_{\rm stag}<1$ usefulness for the estimation of the staggered combination $\theta_1 -\theta_2 + \dots$
The latter is relevant for differential sensing since allowing for common-mode phase noise between neighboring ensembles.

\begin{figure}[t!]
\centering
\includegraphics[width=0.95\columnwidth]{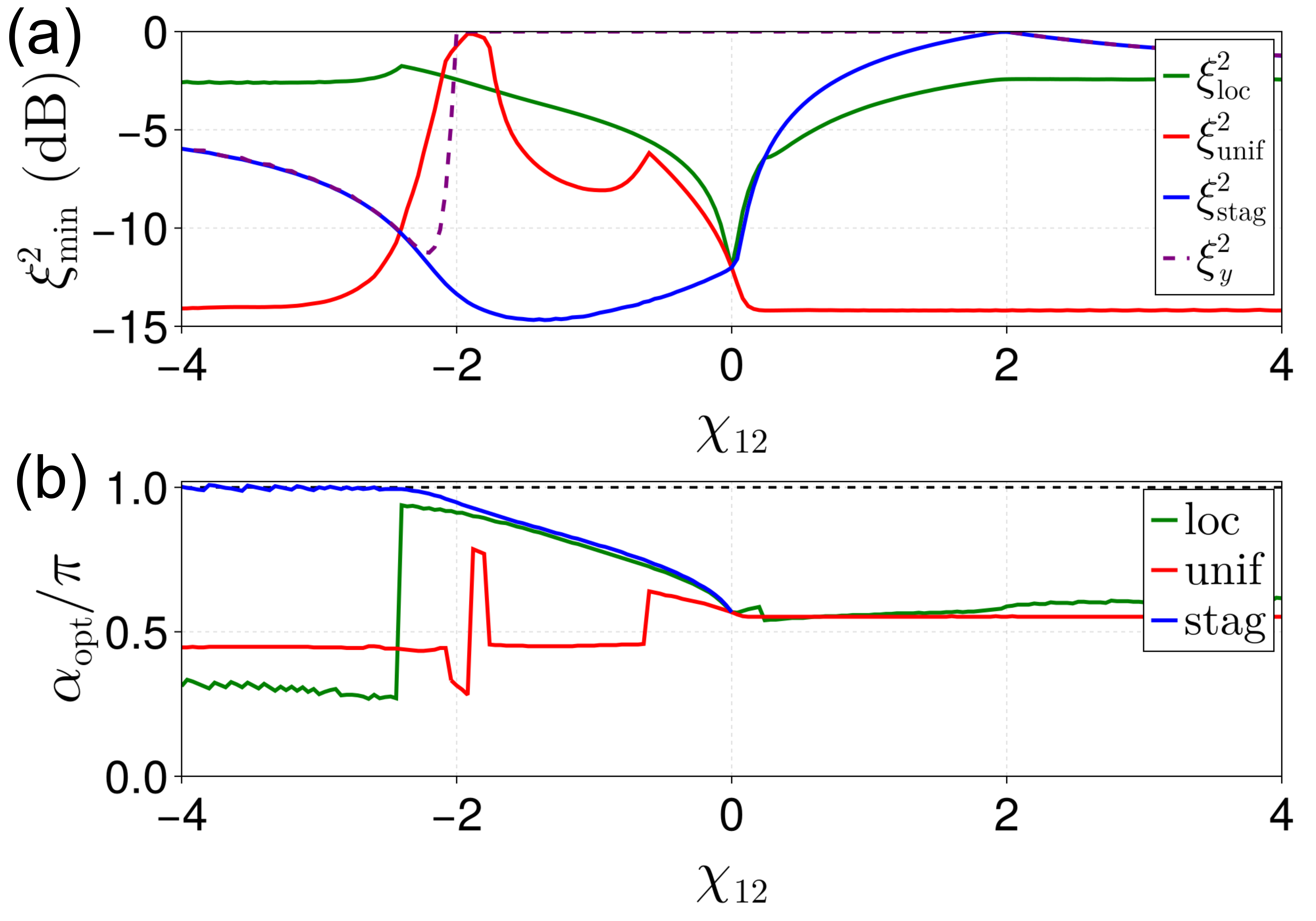}
\caption{(a) Minimal collective (red uniform, blue staggered) and local (green) squeezing achieved in time versus $\chi_{12}$, with $\chi_{loc} = 1$. The dashed purple line corresponds to the staggered squeezing along the fixed $y$ direction. (b) Squeezing angle $\alpha_{\rm opt}$ at the time of the minimum for different values of $\chi_{12}$. The dashed line corresponds to $\alpha = \pi$. Here, we took $N_{1} = N_{2} = 100$. The squeezing angle for uniform squeezing for $\chi_{12}$ around $-1.5$  displays large fluctuations which are due to the very small amount of uniform squeezing for these values of $\chi_{12}$ and are not physically relevant. For similar reasons, we did not show $\alpha_{\rm opt}$ for the staggered squeezing for positive $\chi_{12}$.}
\label{fig2}
\end{figure}

{\it Collective spin squeezing dynamics for $M=2$ ensembles.---} 
We first consider the simplest case $M=2$. 
We initialize each ensemble in a CSS pointing along $x$ axis in the Bloch sphere, and let the system evolve under the unitary dynamics dictated by $\hat{H}$ for a time $t$. 
First, in the case $\chi_{12} = 0$, we simply obtain two independent OAT Hamiltonians~\cite{kitagawa_squeezed_1993}. 
The squeezing matrix is diagonal and we obtain the same amount of squeezing for the local and the collective squeezing parameters. 
The OAT model is analytically solvable and the optimal amount of squeezing is known to scale as $N_{1}^{-2/3}$ \cite{kitagawa_squeezed_1993}. 
%
%
%
A more interesting behavior arises when both $\chi_{loc}$ and $\chi_{12}$ are non-zero, see Fig.~\ref{fig2}.
However, for arbitrary values of $\chi_{\rm loc}$ and $\chi_{12}$, it is not possible to obtain analytical results since the different terms in the Hamiltonian do not commute. 
We focus on two specific but representative values of $\chi_{12}$ for which it is possible to write the Hamiltonian (\ref{hamM}) in terms of collective spin operators and thus derive analytical or semi-analytical results. 

First, for $\chi_{12} = 2 \chi_{loc}$, Eq.~(\ref{hamM}) reduces to
\begin{equation}
    \label{ham_unif}
    \hat{H} = -\chi_{\rm loc} \mathbf{J}_{\rm unif}^{2} +  \chi_{\rm loc} (J^{z}_{\rm unif})^{2}.
\end{equation}
Since $\mathbf{J}_{\rm unif}^{2}$ and $J^{z}_{\rm unif}$ commute and the initial state (product of two CSS) is an eigenstate of $\mathbf{J}_{\rm unif}^{2}$, Eq.~(\ref{ham_unif}) describes an OAT dynamics for the collective spin.
The minimal eigenvalue of the squeezing matrix is obtained for the uniform collective spin, with a scaling of the minimum squeezing as $(2N)^{-2/3}$.
We introduce the squeezing angle $\alpha_{\rm opt}$ in the $yz$ plane (defined $[{\rm mod} ~\pi]$) such that $\alpha_{\rm opt} = 0$ or $\pi$ when $y$ is the most squeezed component and $\alpha_{\rm opt} = \pi/2$ when $z$ is the most squeezed component. With this definition, we obtain a squeezing angle for the uniform squeezing slightly larger than $\pi/2$, with $\alpha_{\rm opt} - \pi/2 \sim 1/(2N)^{1/3}$ in the limit of large $N_{1,2}$ at the time of optimal squeezing \cite{kitagawa_squeezed_1993}.
The maximally squeezed states generated by this dynamics can be visualized by the Wigner distribution in the Bloch sphere with $J_{\rm unif}^{x,y,z}$ axes, see Fig.~\ref{fig1}(b). 
This representation is possible because the uniform collective spin is exactly conserved during the dynamics, which is constrained on the surface of a Bloch sphere of radius $\sqrt{N(N+1)}$.
The dynamics also generates a small amount of local squeezing $\xi^{2}_{\rm loc}$ -- which does not scale with system size and is bounded from below by $\xi^{2}_{\rm loc}> 1/2$ -- and no staggered squeezing $\xi^{2}_{\rm stag}$.
The numerical results of Fig.~\ref{fig2}, reporting the the squeezing parameters (minimized over time) as a function of $\chi_{12}$,   show that this behavior is representative for $\chi_{12}/\chi_{loc} >0$, where we find large uniform squeezing at a tilted angle and low staggered and local squeezing.   

The second interesting choice is $\chi_{12} = -2 \chi_{\rm loc}$.
In this case the Hamiltonian can be written as
\begin{equation}
    \label{ham_stag}
    \hat{H} = \chi_{\rm loc} \mathbf{J}_{\rm unif}^{2} +  \chi_{\rm loc} (J^{z}_{\rm stag})^{2}.
\end{equation}
%
As shown in Fig.~\ref{fig2}(a) it leads to strong staggered squeezing while it suppresses the uniform one.
Yet, the dynamics generated by Eq.~(\ref{ham_stag}) strongly differs from a staggered OAT dynamics due to the competition between $\mathbf{J}_{\rm unif}^{2}$ and $(J^{z}_{\rm stag})^{2}$. 
The optimal squeezing angle $\alpha_{\rm opt}$ is reported in Fig.~\ref{fig2}(b).
Interestingly the angle approaches $\pi$ showing that the state is elongated along the $z$ axis 
This result in particular is relevant for practical metrological application, e.g. the implementation of interferometric measurements with the state - discussed in the last section - as the state would not require an additional rotation before the interferometric sequence, which would make it more robust to preparation noise.

To further investigate the properties of this state, we visualize it in Fig.~\ref{fig1}(b) by looking at the Wigner function on a Bloch sphere of axes $J^x_{\rm unif}$, $J^y_{\rm stag}$ and $J^z_{\rm stag}$.
Notice that up to the time of the optimal squeezing, the staggered collective spin length is almost conserved (see SM C), and we can restrict ourselves to the surface of such Bloch sphere.
Plotting the maximally squeezed states generated by the dynamics at $\chi_{12} = -2 \chi_{\rm loc}$, Fig.~\ref{fig1}(b), we can clearly see that difference between elongation direction with respect to the OAT dynamics at $\chi_{12} = 2 \chi_{\rm loc}$, shown in Fig.~\ref{fig1}(c).
To understand more in details the dynamics, we use a classical model and study the semiclassical trajectories obtained from the equation of motions given in the Appendix (Eq. \ref{class_traj}).
These equations describe the evolution of the state on the Bloch sphere of largest radius for the collective spin ${\bf J}^{2}_{\rm stag/unif}$, see Fig.~\ref{fig1} (b-c). For the OAT Hamiltonian the trajectories are circles of fixed height (rotating counter-clockwise in the Northern hemisphere and clockwise in the Southern hemisphere). However, for $\hat{H}$ at $\chi_{12} = -2 \chi_{loc}$, we get a line of fixed points (corresponding to $J^{y}_{\rm stag} = 0, ~J^{x}_{\rm unif} >0$, while trajectories circle around the point $J^{x}_{\rm unif} = - 2S$ (see Appendix and SM D). This explains why we observe a squeezing of the Wigner distribution along the $y$ direction.
This evolution is qualitatively similar to the one obtained with a two-axis counter-twisting dynamics \cite{kitagawa_squeezed_1993,KajtochPRA2015,Luo2025}, but the structure of the fixed points is rather different.

The above discussion, restricted to $\chi_{12}/\chi_{loc} = -2$, is representative for $\chi_{12}/\chi_{loc}$ negative but not too largely negative, as shown in Fig.~\ref{fig2}.
In this regime we observe staggered spin squeezing for the $y$ component: we find $\alpha_{\rm opt} = \pi$ for $\chi_{12} < -2.4$.
This observation is also consistent with the fact that we observe non-zero squeezing at $\alpha = 0$ (dashed purple line in  Fig.~\ref{fig2} (a)) only for $\chi_{12} < -2$, while the standard OAT dynamics is unable to generate squeezing along $y$.

\begin{figure}[t!]
\centering
\includegraphics[width=0.9\columnwidth]{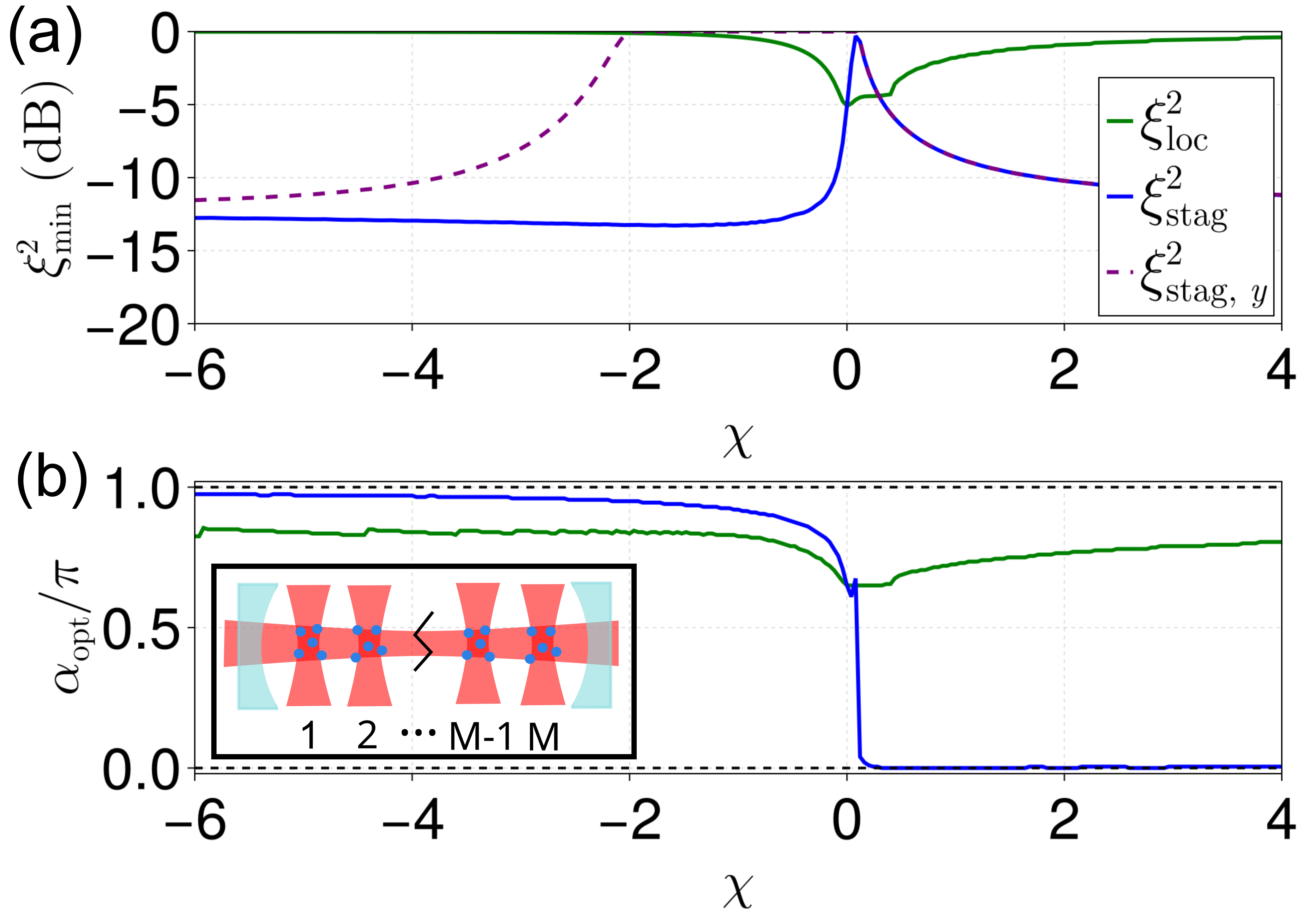}
\caption{ (a) Minimal collective staggered (blue) and local (green) squeezing achieved in time for different values of $\chi$ with $\chi_{loc} = 1$. The dashed purple line corresponds to the staggered squeezing along the fixed $y$ direction. (b) Squeezing angle $\alpha_{\rm opt}$ at the time of the minimum for different values of $\chi$. In both figures (a, b) we took $N = 10$ and $M=20$. The black dashed lines correspond to $0$ and $\pi$.
Inset: schematic of the system with $M$ tweezers.
}
\label{fig3}
\end{figure}

{\it Generalization to M clouds.---} We now discuss the case of an arbitrary number of clouds $M$. 
The case of large $M$ is crucial for distributed sensing in general, but even more so in our case which is currently limited to a small number of atoms within each tweezer to avoid atom losses by direct interactions between the atoms. Therefore increasing the metrologic sensitivity of this platform goes through increasing the number of tweezers instead of increasing the number of atoms in each tweezer.
%
%
Since we are interested in differential measurement, we will focus on the dynamics leading to staggered squeezing: we will consider even $M$ and $\chi_{ij}=(-1)^{i+j-1} \chi$ in Eq.~(\ref{hamM}).
In this case, for $\chi = -2$, we can again exactly rewrite the Hamiltonian in a collective form analogous to Eq.~\ref{ham_stag} up to constant terms. 
The whole dynamics is equivalent to the dynamics of only two subsystems, and is fully determined by the spin length of each subsystem $MN/2$.
For any other value of $\chi$, computing exactly the time evolution of such system in intractable with exact diagonilization methods for $M>6$ for $N=10$. In order to increase the cloud number, we thus rely on an approximate methods, based on truncated cumulant expansion (TCE) of the spin observables, as introduced in \cite{Trifa24}. 
This approach consist in approximating high order correlation functions for the spin moments by decomposing them into lower order correlation functions, assuming that all cumulants higher than a given order $n$ (here $n=2$) vanish. This leads to a polynomial set of classically coupled non-linear differential equations that we can solve numerically (see SM E).
With this method, it is easy to increase the number of clouds, while we are limited to rather small ensembles in each cloud ($N \sim 10)$.
We show in Fig.~\ref{fig3} the optimal squeezing obtained in time for an ensemble of $M=20$ clouds with each $N=10$ atoms. Again, we observe a large amount of staggered squeezing around $\chi = -2$, while there is almost no local squeezing, apart from a small region for $\chi$ slightly positive.
We also find that for large negative $\chi$ of for any strictly positive $\chi$, we have a large amount of staggered squeezing along the $y$ direction, without requiring any extra rotation.

Fig \ref{fig3} (b) shows the optimal squeezing angle versus $\chi$, and we see that this angle goes to $0$ for positive $\chi$ for the staggered squeezing, which is compatible with the previous observation. Note that in both (a) and (b) we chose not to show the results for the uniform squeezing, as we either found small to no amount of squeezing or unstable TCE simulation results depending on the value of $\chi$ (see SM F).

%
 
{\it Differential Interferometry.---}
Finally, we investigate application to differential interferometry. 
We consider the case of even values of $M$ and assume that the subsystem $j$ acquires a phase $\theta_{A}$ if $j$ is odd and $\theta_{B}$ if $j$ is even.
Phase accumulation is described by the transformation $e^{-i \theta_j S_j^z}$, modeling a Ramsey interferometer and applied simultaneously to each ensemble.
We want to show that preparation of multiple ensembles with collective spin squeezing is particularly useful for measuring the phase difference $\delta \theta =  \theta_{A} - \theta_{B}$ with high precision when both are subject to random common noise.
The scheme can be also applied for the synchronization of two atom clocks, in which case $\theta_{A,B}$ is replaced by $\nu_{A,B}$, the clock transition frequency.
Here we consider the case of a large noise environment, with phase noise uniformly distributed between $0$ and $2 \pi$, which is the worst case scenario compared to the case of small correlated noise, as studied in \cite{malia_nat_2022}. 

\begin{figure}[t!]
\centering
\includegraphics[width=0.95 \columnwidth]{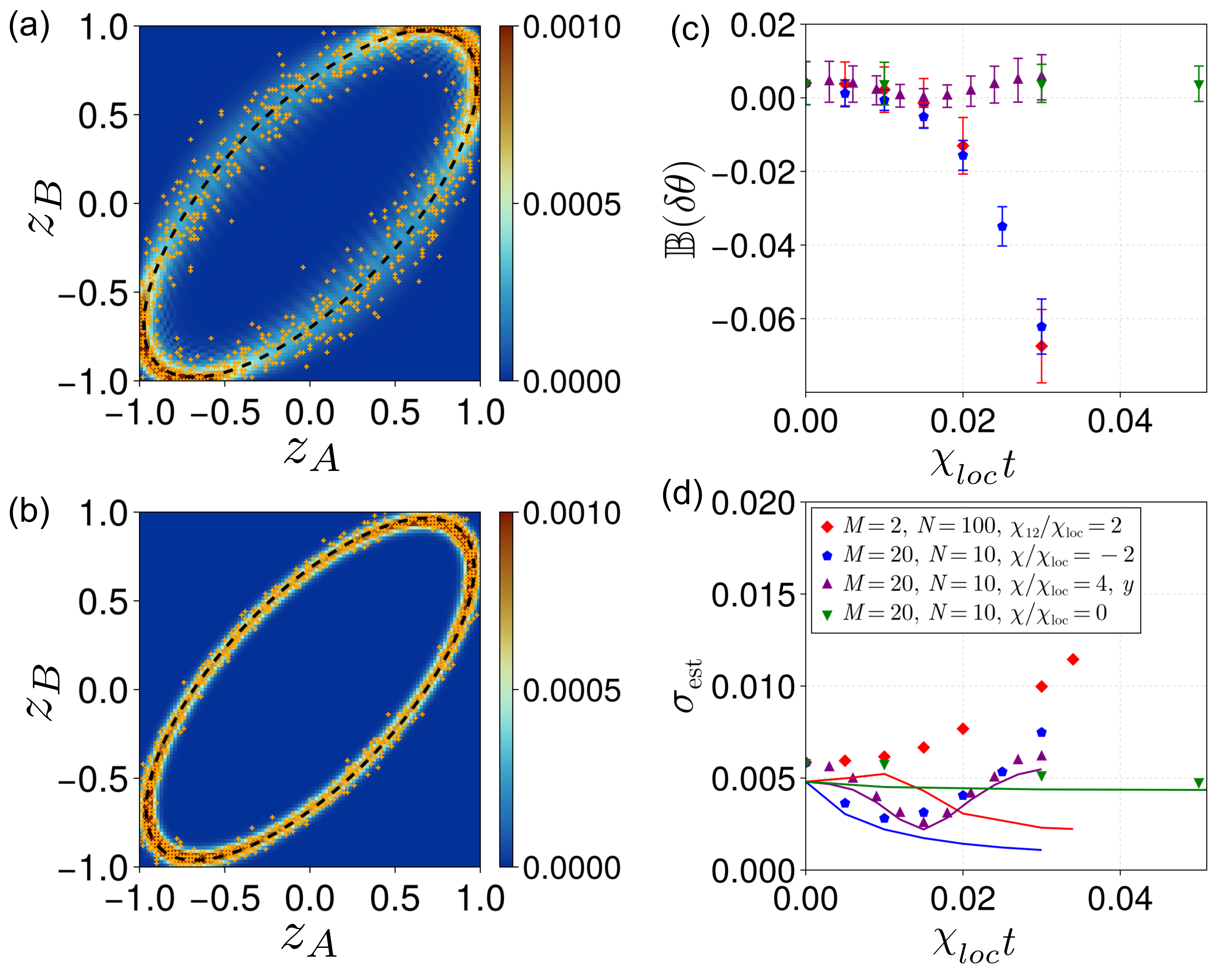}
\caption{ Ellipse fitting from $1000$ points sampled from the distribution $\bar{\cal P} \left( m_{z}^{A}, m_{z}^{B}, \chi_{loc}t=0.01, \pi/4 \right)$ for (a) a collective uniform OAT dynamics ($\chi_{loc} = 1, \chi_{12} = 2$, $M=2$ and $N =100$) and (b) a staggered squeezed state ($M=20$, $N=10$, $\chi/\chi_{\rm loc}  =  4$, without local rotation to change the squeezing direction). We note $z_{j} = 4m_{z}^{j}/(MN)$, $j = A, B$. The black dashed ellipse is obtained with a non linear geometric fit. (c) Time evolution of the bias and (d) standard deviation of the estimator for the different Hamiltonian parameters. The solid lines correspond to the Cramèr-Rao bound, computed from the classical Fisher information of the density probability $\bar{\cal P}$.}
\label{fig4}
\end{figure}

We first numerically compute the joint probability $\bar{\cal P} ( m_{z}^{A}, m_{z}^{B} \vert t,\delta \theta)$ of measuring $m_{z}^{A}$ and $m_{z}^{B}$, the measurement outcomes of $J^{z}_{A} \equiv \sum_{j~{\rm odd}} S^{z}_{j}$ and $J^{z}_{B}\equiv \sum_{j~{\rm even}} S^{z}_{j}$, after a time evolution of duration $t$ and averaging over the phase noise $\epsilon$ and a final $\pi/2$ pulse around $J^{y}_{\rm unif}$ (see Appendix). 
Accessing the exact probability distribution $\bar{\cal P}$ requires access to the full wavefunction for the total system, which limits us to relatively small system sizes (up to $N=100$ for $M=2$ using ED). In order to realize the simulation with large number of tweezers (up to $M=20$ and $N=10$ to have the same total metrological resource), we used both TCE simulations and gaussian approximation of the probability distribution (see SM G). We built our estimator of the phase difference $\delta \theta$ using a geometric method for ellipse fitting \cite{GanderBIT1994, FosterOL02}, using $\eta_{ell} = 1000$ points sampled from the distribution $\bar{\cal P}$ (see Fig.~\ref{fig4} (a,b)). We then extract the bias and variance of the estimator from $\eta = 1000$ results of the ellipse fitting protocol for different values of the interaction strength $\chi$ and for different times.

We show our results in Fig.~\ref{fig4} (c,d). First, we do not see any improvement in sensitivity compared to the CSS when using the uniform squeezing - as obtained with a OAT dynamics with $M=2$, $N=100$ and $\chi = 2$. Improving the sensitivity of the differential interferometer requires squeezing of the staggered component $J^{y}_{\rm stag}$ and not the uniform one $J^{y}_{\rm unif}$. 
Indeed, we observe a strong improvement in sensitivity when using staggered squeezed states. Interestingly, the optimal sensitivity (i.e. smallest $\sigma$) is reached before the time of maximal squeezing - e.g. $\chi_{loc} t_{\rm opt} \approx 0.03$ for $\chi$ = -2 - at a time $\chi_{loc} t^{*} \approx 0.01$ for the same value of $\chi$. We also show that the bias, originally compatible with $0$, increases after $t^{*}$. This confirms the results of Ref.~\cite{CorgierQST2025} (using only local squeezing) pointing to an optimal amount of squeezing for the ellipse fitting protocol. Using the entanglement between the atom clouds, we can reach an even greater sensitivity compared to the uncorrelated case, using the staggered squeezing ($\chi_{loc} = 1, \chi = -2$ or $4$). The time $t^{*}$ is even lower in the staggered case compared to the local case, which is a second advantage of using non-local squeezing compared to local squeezing only. 
Finally, we compare the uncertainty of our estimator with the Cramèr-Rao bound, obtained from the classical Fisher information of $\bar{\cal P}$. Fig.~\ref{fig4} (d) reveals that for the staggered squeezing our estimator almost saturates this bound. This also shows that we are well below the SQL, by almost a factor $2$ for $M=20$ and $N=10$, which could be further improved by considering larger ensembles.

{\it Conclusions.---}
Our results suggest a mode-engineering strategy for distributed quantum sensing in which the interaction graph is designed around the sensing task.
The cavity-coupling matrix then becomes a programmable resource: by matching its eigenmodes to the spatial structure of the target field, future sensor arrays could selectively enhance sensitivity to gradients, spatial curvatures, or Fourier components while remaining insensitive to dominant common-mode fluctuations.
More complex coupling graphs could also generate correlations across several spatial modes, opening the way to multiparameter field imaging associated with gradients and spatial
curvatures, and allowing the same sensor array to be reconfigured for different observables without changing its physical architecture.
A key next step is to determine how this mode engineering performs under
realistic conditions, including photon loss, spontaneous emission,
inhomogeneous atom--cavity coupling, and atom-number fluctuations, each requiring a dedicated analysis not addressed in this work.
The rapid preparation times enabled by coherent cavity dynamics are encouraging in this regard.
Moreover, the restriction of this study to squeezed states instead of Heisenberg-limited states - which are known for their fragility in spite of their unmatched metrological advantage - also offers a strong advantage in terms of robustness versus different types of noise.

\begin{acknowledgments}
{\it Acknowledgments.---} We acknowledge discussions with J. Reichel, P. Treutlein, L. Salvi, A. Sinatra, and G.M. Tino.
This publication has received funding under Horizon Europe programme HORIZON-CL4-2022-QUANTUM-02-SGA via the project 101113690 (PASQuanS2.1). 
LP acknowledges financial support by the QuantEra project SQUEIS.
\end{acknowledgments}




\begin{center}
    {\bf APPENDIX}
\end{center}

{\it Multiparameter squeezing} The squeezing matrix is a $M \times M$ matrix ($M$ being the number of subsystems) defined as \cite{gessner_multiparameter_2020}:
\begin{equation} \label{squeezematrix}
    \Xi^{2} \left( \theta_{1}, \theta_{2} \cdots \theta_{M} \right) \equiv \left( \xi_{ij}^{2} \right)_{ij =1,M}
\end{equation}
with the diagonal elements being the local Wineland squeezing parameter for a fixed $\theta_{j}$:
\begin{equation}
    \xi^{2}_{jj} \left( \theta_{j} \right) = \frac{ N_{j}  \mathrm{Var} \left( \cos \left( \theta_{j} \right)  S^{y}_{j} + \sin \left( \theta _{j} \right) S^{z}_{j}  \right)}{\vert \langle S^{x}_{j} \rangle \vert ^{2}}
\end{equation}
while the off-diagonal elements ($\i \neq j$) are sensitive to correlation between sub ensembles and are given by:
\begin{align} \label{xi12}
\xi^{2}_{ij} & \equiv  \xi^{2}_{ij} \left( \theta_{i}, \theta_{j} \right) = \xi^{2}_{ji} \left( \theta_{i}, \theta_{j} \right) \nonumber \\
& \equiv  \frac{\sqrt{ N_{i} N_{j} } \langle \left[ c_{i} S^{y}_{i} + s_{i} S^{z}_{i} \right]  \left[ c_{j}  S^{y}_{j}  + s_{j} S^{z}_{j} \right]\rangle }{\vert \langle  S^{x}_{i} \rangle \langle S^{x}_{j} \rangle \vert}
\end{align}
with $c_{k} = \cos \left( \theta_{k} \right)$ and $s_{k} = \sin \left( \theta_{k} \right)$. The numerator corresponds to the covariance of the rotated spin component between the two susbsystems $i$ and $j$, with the assumption that the mean value of each spin component is $0$ (since they are orthogonal to the polarization direction $x$).
Introducing the vectors $\nu_{\rm unif} = 1/\sqrt{M}\begin{pmatrix}
    1& \cdots & 1
\end{pmatrix} ^{T}$ and $\nu_{\rm stag} = 1/\sqrt{M} \begin{pmatrix}
    1 & -1 & \cdots 1 & -1
\end{pmatrix} ^{T}$, we can define the collective multiparameter squeezing as the minimum over all possible angle $\theta_{j}$ of the average value of the squeezing matrix:
\begin{equation}
    \xi^{2}_{\rm col} = \min_{\theta_{1}, \cdots \theta_{M}} \nu^{T}_{\rm col} \Xi^{2} \left( \theta_{1}, \cdots \theta_{M} \right) \nu_{\rm col}
\end{equation}
with $\rm col$ being either $\rm unif$ for the uniform case (corresponding to the squeezing of the sum) or $\rm stag$ for the staggered case (corresponding to the squeezing of the difference). Finally, we can extract the local squeezing (equivalent to the single spin Wineland parameter) as 
\begin{equation}
    \xi^{2}_{\rm loc} = \min_{\theta_{1}} \xi^{2}_{11} = \min_{\theta_{j}}  \xi^{2}_{jj} ~\forall j
\end{equation}
assuming all subsystems have the same atom number and are invariant under permutation.

\vspace{2mm}

{\it Semiclassical trajectories} To understand more in details the dynamics, we rely on a classical model, and study the semiclassical trajectories.
We first introduce two classical variables for the state of each subsystem ($z_{i}$, $\phi_{i}$), such that the state of each subsystem is given by $ \mathbf{S}_{i} = \Big( \sqrt{1-z_{i}^{2}} \cos \left( \phi_{i} \right), ~ \sqrt{1-z_{i}^{2}} \sin \left( \phi_{i} \right), ~  z_{i} \Big)$. It is then possible to write the classical equations of motions (under some assumptions described in SM D) of the full system only in terms of $\delta z = z_{1} - z_{2}$ and $\delta \phi = \phi_{1} - \phi_{2}$, and we obtain the following set of equations:
\begin{equation}
\label{class_traj}
    \frac{{\rm d} \delta z}{{\rm d} \tau} =  2 \left( 1 - \frac{\delta z^{2}}{4} \right) \sin \left( \delta \phi \right) \quad
    \frac{{\rm d} \delta \phi }{{\rm d} \tau} = 2 \left( 1 - \cos \left( \delta \phi \right) \right) \delta z 
\end{equation}
with $\tau$ a dimensionless parameter for the evolution of the trajectories. These equations describe the evolution of the state on the Bloch sphere associated with the largest eigenvalue subspace of the staggered collective spin, $\mathbf{J}_{\rm stag}^{2}$. 
On the other hand, for a OAT Hamiltonian, the equations of motion are given by:
\begin{equation}
\label{class_trajOAT}
    \frac{{\rm d}  Z}{{\rm d} \tau} =  0 \quad
    \frac{{\rm d} \Phi }{{\rm d} \tau} =  Z 
\end{equation}
where $Z = z_{1} + z_{2}$ and $\Phi = \phi_{1} + \phi_{2}$, and the evolution is now on the Bloch sphere associated with the largest eigenvalue subspace of the uniform collective spin, $\mathbf{J}_{\rm unif}^{2}$. 
For the OAT Hamiltonian the trajectories are simply circles of fixed height ($Z = 0$ (rotating counter-clockwise in the Northern hemisphere and clockwise in the Southern hemisphere). However, for the cavity Hamiltonian at $\chi_{12} = -2 \chi_{loc}$, we get a line of fixed points at $\delta \phi = 0$, while trajectories circle around $\delta \phi = \pi$ (see SM D for more details). This explains why we observe a squeezing of the Wigner distribution along the $y$ direction, while it expands along the $z$ direction.

{\it Joint probability distribution.---}
We define the joint probability to measure $m_{z}^{B}$  and $ m_{z}^{B}$ as measurement outcome of the measurement of $J^{z}_{A}$ and $J^{z}_{B}$, for a fixed phase difference $\delta \theta$ and for a realization of the common noise $\epsilon$: 
\begin{align}
{\small}
{\cal P} \left( m_{z}^{A}, m_{z}^{B}, t, \delta \theta, \epsilon \right) &= \vert \langle m_{z}^{A}, m_{z}^{B} \vert e^{i (J^{y}_{A} + J^{y}_{B}) \pi/2} e^{i \delta \theta (J^{z}_{A} - J^{z}_{B})} \nonumber e^{i \epsilon (J^{z}_{A} + J^{z}_{B})} \vert \psi \left( t \right) \rangle \vert ^{2}
\end{align}
with $\vert \psi \left (t \right) \rangle$ the state of the system after time evolution with Hamiltonian $\hat{H}$ for a time $t$ and $\vert m_{z}^{A}, m_{z}^{B} \rangle$ is a product state of two Dicke states along $z$, i.e. two eigenvectors of $J^{z}_{A}$ (resp. $J^{z}_{B}$) with eigenvalues $m_{z}^{A}$ (resp. $m_{z}^{B}$). In the case of only $M=2$, this probability can be reconstructed exactly from the wavevector obtained with exact diagonalization up to $N=100$. However, for $M\leq6$, even having access to the local properties of each ensemble with TCE simulations, we need to approximate the probability for each value of the noise by a multivariate normal law in order to be able to compute the full probability distribution for the collective spin in each subsystem A and B, see SM G for more details.
Finally, we average this probability over the noise distribution probability to obtain the average distribution:
\begin{align}
\bar{\cal P} \left( m_{z}^{A}, m_{z}^{B}, t, \delta \theta \right) &= \frac{1}{2 \pi} \int_{0}^{2 \pi} \mathrm{d} \epsilon \,  {\cal P} \left( m_{z}^{A}, m_{z}^{B}, t, \delta \theta, \epsilon \right).
\end{align}


\onecolumngrid

\begin{center}
    Supplemental Material

\end{center}

\section*{A- Derivation of $H_{cav}$}

\begin{figure}[ht!]
\includegraphics[width=0.4\textwidth]{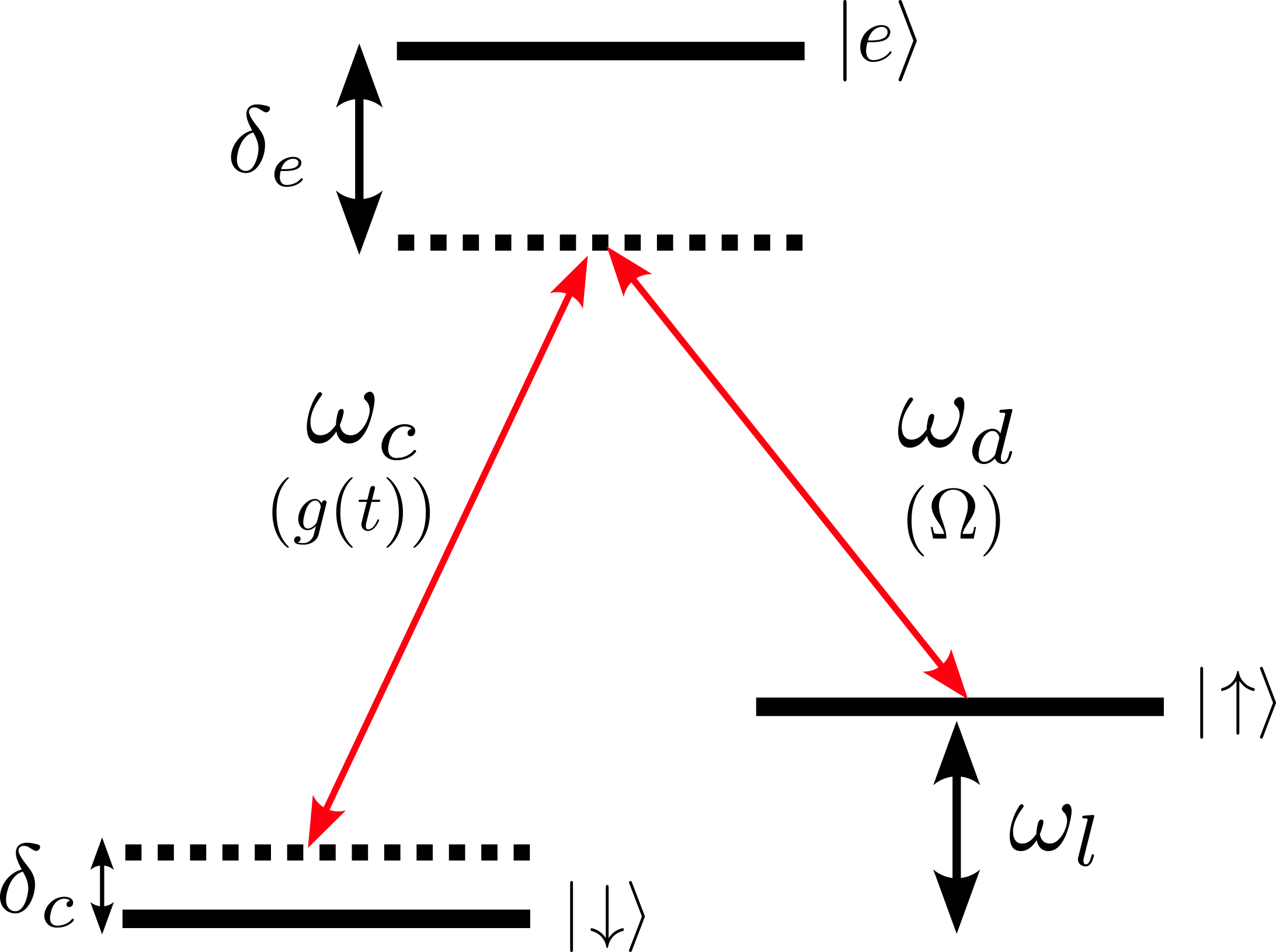}
\caption{ Schematic of the three levels and atomic transitions used to engineer the cavity Hamiltonian.
}
\label{figA}
\end{figure}

Following the work of Ref.~\cite{Chu2026}, we start with a three-level system for each atom cloud, with two states $\vert \uparrow \rangle$ and $\vert \downarrow \rangle$ from the ground state manifold of the atoms, and an excited state $\vert e \rangle$. The states $\vert \uparrow \rangle$ and $\vert \downarrow \rangle$ are coupled via a two-photon process involving the excited state $\vert e \rangle$ with the cavity laser (resonant at $\omega_{c}$) and a Rabi drive (resonant at $\omega_{d}$, with Rabi frequency $\Omega$). We note $\delta_{e} = \omega_{d} - \omega_{e}$ the detuning between the transition between $\vert \uparrow \rangle$ and $\vert e \rangle$ and the Rabi drive, and $\delta_{c} = \omega_{d} + \omega_{0} - \omega_{c}$ the detuning between the transition between $\vert \uparrow \rangle$ and $\vert \downarrow \rangle$ and the two-photon Raman process (see Fig. \ref{figA}).

Adding a magnetic field gradient along the cavity axis, it is possible to tune the transition between $\vert \uparrow \rangle$ and $\vert \downarrow \rangle$ to obtain a position-dependent resonant frequency, $\omega_{l} = \omega_{0} + \Delta_{z} (l-1)$.

Assuming that the detuning $\delta_{e} \gg \Omega$, and that the population in the excited state is small at the initial time, we can adiabatically eliminate the excited states from the Hamiltonian, and obtain a first effective two-mode model coupled to the cavity mode:

\begin{equation}
    H = \omega_{c} a^{\dagger} a + \sum_{l} \omega_{l} S^{z}_{l} +  \sum_{l} \frac{\Omega g}{\delta_{e}} \left[ a S^{+}_{l} e^{-i\omega_{d} t} + a^{\dagger} S^{-}_{l} e^{i \omega_{d} t} \right]
\end{equation}

with $a$ the bosonic annihilation operator for the cavity mode, $2g$ is the single-atom vacuum Rabi splitting of the cavity and $S_{l}^{\alpha}$ is the collective spin operator at site $l \in \left[1, M\right]$:
\begin{equation}
    S^{\alpha}_{l} = \sum_{i = 1}^{N_{l}} \sigma_{i}^{\alpha}.
\end{equation}

Moving to the rotating frame corresponding to the bare cavity mode and Zeeman splitting of the atoms, with a unitary transform given by $U = e^{-i (\omega_{c} a^{\dagger} a + \sum_{l} \omega_{l} S^{z}_{l})t }$  we obtain the following Hamiltonian:
\begin{equation}
    H_{\rm RF} = - \delta_{c} a^{\dagger} a +  \sum_{l} \frac{\Omega g}{\delta_{e}} \left[ a S^{+}_{l} e^{-i (l-1) \Delta_{z} t} + a^{\dagger} S^{-}_{l} e^{i (l-1) \Delta_{z} t} \right]
\end{equation}

Finally, assuming that we have a far detuned cavity mode, $\delta_{c} \gg \frac{\Omega g}{\sqrt{S} \delta_{e}}$, we can adiabatically eliminate the cavity mode from the Hamiltonian (by setting $\frac{{\rm d} a}{{\rm d} t} = 0$), and we obtain the following effective spin-spin interacting Hamiltonian:

\begin{equation}
    H_{eff} = \frac{1}{2} \sum_{l, m} \frac{1}{\delta_{c}} \left( \frac{\Omega g}{\delta_{e}} \right)^{2} \left[   S^{+}_{l} S^{-}_{m} e^{- i (l-m) \Delta_{z} t} + S^{-}_{l} S^{+}_{m} e^{+ i (l-m) \Delta_{z} t} \right]
\end{equation}

In the case of strong gradient $\Delta_{z} \gg \frac{1}{\delta_{c}} \left( \frac{\Omega g}{\delta_{e}} \right)^{2} $, we can perform a rotating wave approximation, and we simply obtain local quadratic Zeeman terms:
\begin{equation}
    H_{eff} = \frac{1}{2} \sum_{l} \frac{1}{\delta_{c}} \left( \frac{\Omega g}{\delta_{e}} \right)^{2}  \left[ S^{+}_{l} S^{-}_{l} + S^{-}_{l} S^{+}_{l} \right] = - \sum_{l} \frac{1}{\delta_{c}} \left( \frac{\Omega g}{\delta_{e}} \right)^{2} \left( S^{z}_{l} \right)^{2}
\end{equation}

In order to obtain non vanishing long-range coupling between the clouds, we follow the same path as Ref.~\cite{Periwal2021}, introducing a temporal modulation of the local couplings $g \equiv g (t)$. We note $\Tilde{\chi} \left( t \right) = \frac{1}{\delta_{c}} \left( \frac{\Omega g(t)}{\delta_{e}} \right)^{2} $.
Then, any given set of effective coupling between neighboring sites at distance r, $\chi \left( r \right)$, can be obtained via setting $\Tilde{\chi} (t) = \sum_{r} e^{-i r \Delta_{z} t } \chi(r)$, by time averaging the Floquet Hamiltonian in the rotating Frame, which leads to:
\begin{equation}
    H_{eff} = \frac{1}{2} \sum_{l, m} \chi_{lm} \left[ S^{+}_{l} S^{-}_{m} + S^{-}_{l} S^{+}_{m} \right]
\end{equation}
where we introduced the notation $\chi_{lm} = \chi(l-m)$. Finally, to obtain our Hamiltonian $H_{cav}$, we first set $\chi_{lm} = \chi$  to have a constant interaction. Then, we add an extra drive field, without modulation and with opposite sign detuning for $\delta_{c}$, in order to adjust the ratio between the local quadratic terms and interaction term, as suggested in Ref.~\cite{Chu2026}. 

\section*{B- Time evolution of the Multiparameter squeezing for $M=2$}
\label{app_ham_cav}

\begin{figure}[htbp!]
\includegraphics[width=0.99\textwidth]{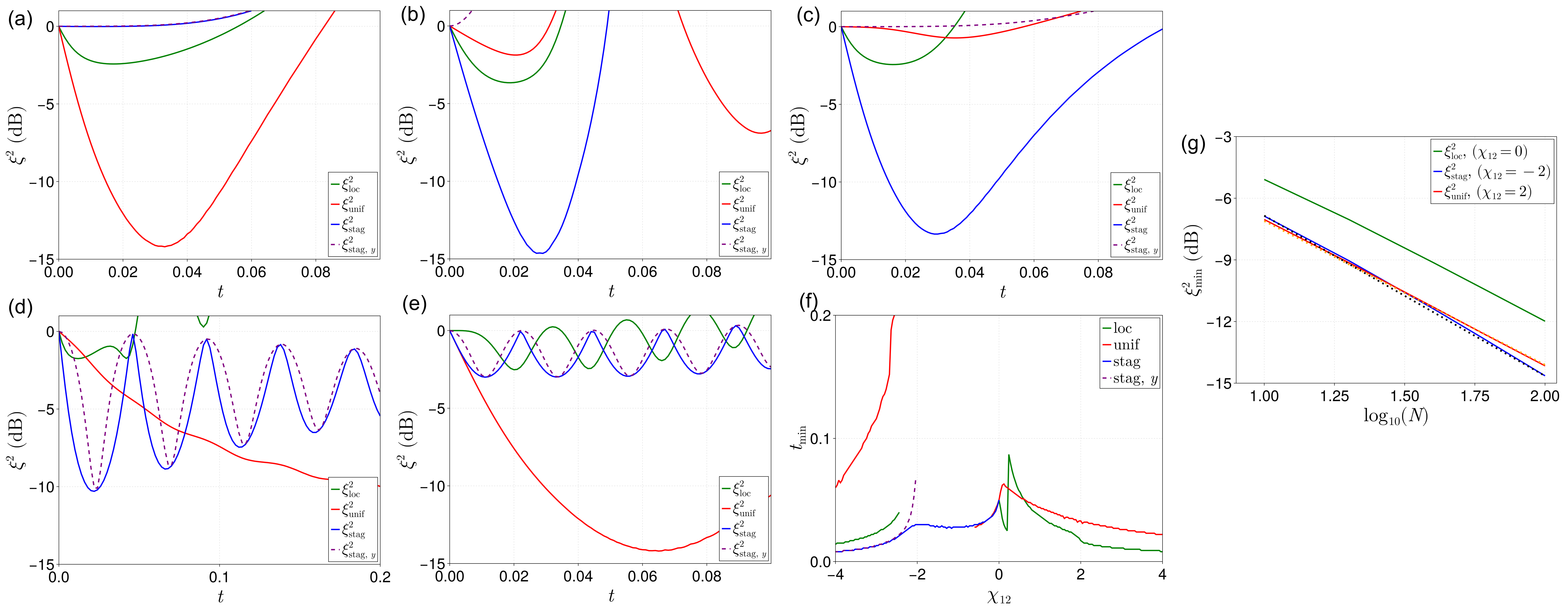}
\caption{ (a-e) Time evolution of the eigenvalues of multiparameter squeezing matrix (in dB), with $\hat{H}$ for (a) $\chi_{12} = 2,  ~\chi_{loc} = 1 $; (b) $\chi_{12} = -1.4,  ~\chi_{loc} = 1 $; (c) $\chi_{12} = -2,  ~\chi_{loc} = 1 $; (d) $\chi_{12} = -2.2,  ~\chi_{loc} = 1 $; (e) $\chi_{12} = -2,  ~\chi_{loc} = 0 $. We show in green solid lines the local squeezing, in solid red the uniform one, in solid blue the staggered one and in dashed purple the staggered one at fixed angle (along $y$). For figures (a) to (d), $t$ is in $1/\chi_{\rm loc}$ unit, while it is in $2/\chi_{12}$ unit for figure (e). (f) Time of the minimum of the eigenvalues of the squeezing matrix versus $\chi_{12}$, with $\chi_{loc} = 1$. For all figures we have taken $M=2$ and $N = 100$. As we did in the main text for Fig.~\ref{fig2} (b), we chose not to plot the time of the minimum in the case of the staggered squeezing for positive $\chi_{12}$ and the uniform squeezing for $\chi_{12}$ around $-2$. (g) Scaling of the minimum of the squeezing parameter with the system size $N$.}
\label{figB}
\end{figure}

Here, we give more details on the time evolution with $\hat{H}$ of the different eigenvalues of the multiparameter squeezing matrix (corresponding to the uniform and staggered collective squeezing)  for various values of $\chi_{\rm loc}$ and $\chi_{12}$. In particular, we will discuss the role of $\chi_{loc}$ and $\chi_{12}$ independently. First, for $\chi_{12} = 0$ and any value of $\chi_{\rm loc}$, the two subsystems are independent and the collective uniform and staggered squeezing are equal to the local squeezing, which obeys a OAT dynamics. Then, for $\chi_{12} = 2$ and $\chi_{\rm loc} = 1$, as shown in the main text, the Hamiltonian obeys a global one axis twisting dynamics, as shown in Fig.~\ref{figB} (a). This leads to the development of strong collective uniform squeezing (red), a small amount of local squeezing (green) but no staggered squeezing (blue and purple). On the other hand, for negative $\chi_{12}$ and $\chi_{\rm loc} = 1$, we have strong staggered squeezing while no uniform squeezing, as shown in Fig.~\ref{figB} (b-d). In particular, for $\chi_{12} = -1.4$ we have more squeezing than what is achievable with a standard OAT dynamics, and this squeezing is reach at a shorter time (see Fig.~\ref{figB} (f) for comparison of the time of minimum). This is also true for $\chi_{12} = -2$, this specific value being of interest as this dynamics is also generalizable to larger even number of ensembles $M>2$. On the other hand, for $\chi_{12} = -2.2$, we have less staggered squeezing than the two previous cases, but this time it is possible to have a large amount of squeezing ready to use without any local rotation of the spin state, see the dash purple line in Fig.~\ref{figB} (d). Finally, for $\chi_{loc} = 0$ (see Fig.~\ref{figB} (e)), we have strong uniform squeezing, and small or no local and staggered squeezing.


Regarding the time of the minimum of the squeezing parameter, it is worth noting that for staggered squeezing the minimum is reached faster for any negative $\chi$ compared to a OAT dynamics ($\chi_{12} = 2$). It is possible to accelerate the uniform squeezing dynamics while maintaining a large amount of uniform squeezing by increasing the value of $\chi_{12}$, with a limit for $\chi_{12} \longrightarrow \infty$ corresponding to the $\chi_{\rm loc} = 0$ case, see Fig.~\ref{figB} (e). 

Finally, we extracted the scaling exponent $\nu$ of the optimal squeezing parameter with system size $N$ at $\chi_{12} = -2 \chi_{\rm loc}$, and compared this scaling with the one obtained for a OAT dynamics. The results shown in Fig.~\ref{figB} (g) indicate that the scaling with $N$ is improved for the ACT dynamics ($\nu \approx -0.78$) with respect to the OAT dynamics ($\nu = -2/3$).

\section{C- Wigner distribution of the evolved state}
\label{app_wigner}


The Wigner distribution ${\cal W}$ of a quantum state $\vert \Psi \rangle$, corresponding to a spin length $S$, associated to the density matrix $\rho = \vert \Psi \rangle \langle \Psi \vert$ is defined as:
\begin{equation}
    {\cal W} \left( \theta, \phi \right) = \mathrm{Tr} \left( \rho \Delta_{S} \left( \theta, \phi \right) \right)
\end{equation}
with $\left( \theta, \phi \right)$ the polar and azimuthal angle on the Bloch sphere of radius $\sqrt{S(S+1)}$, and $\Delta_{S}$ defined as \cite{varilly_moyal_1989, klimov_generalized_2017}:
\begin{equation}
    \Delta _{S} \left( \theta, \phi \right) = \sqrt{\frac{4 \pi}{2S +1}} \sum_{L=0}^{2S} \sum_{M = -L}^{L} Y^{*}_{LM} \left( \theta, \phi \right) T^{(S)}_{LM}, \quad T^{(S)}_{LM} = \sqrt{\frac{2l+1}{2S+1}} \sum_{m_{1}, m_{2} = -S}^{S} \begin{pmatrix} S & S & L \\ m_{1} & m_{2} & M    \end{pmatrix} \vert S, m_{1} \rangle \langle S, m_{2} \vert
\end{equation}
with $Y_{LM}$ the usual spherical Harmonics, $T^{(S)}_{LM}$ the spherical tensor operators, $\begin{pmatrix} S & S & L \\ m_{1} & m_{2} & M    \end{pmatrix}$ the $3-j$ Wigner symbol (related to the Clebsh-Gordan coefficients) and $\vert S, m \rangle$ the so-called Dicke states, satisfying $\mathbf{S}^{2} \vert S, m \rangle = S(S+1) \vert S, m \rangle$ and $S^{z} \vert S, m \rangle = m \vert S, m \rangle $.

\begin{figure}[ht!]
\includegraphics[width=0.9\textwidth]{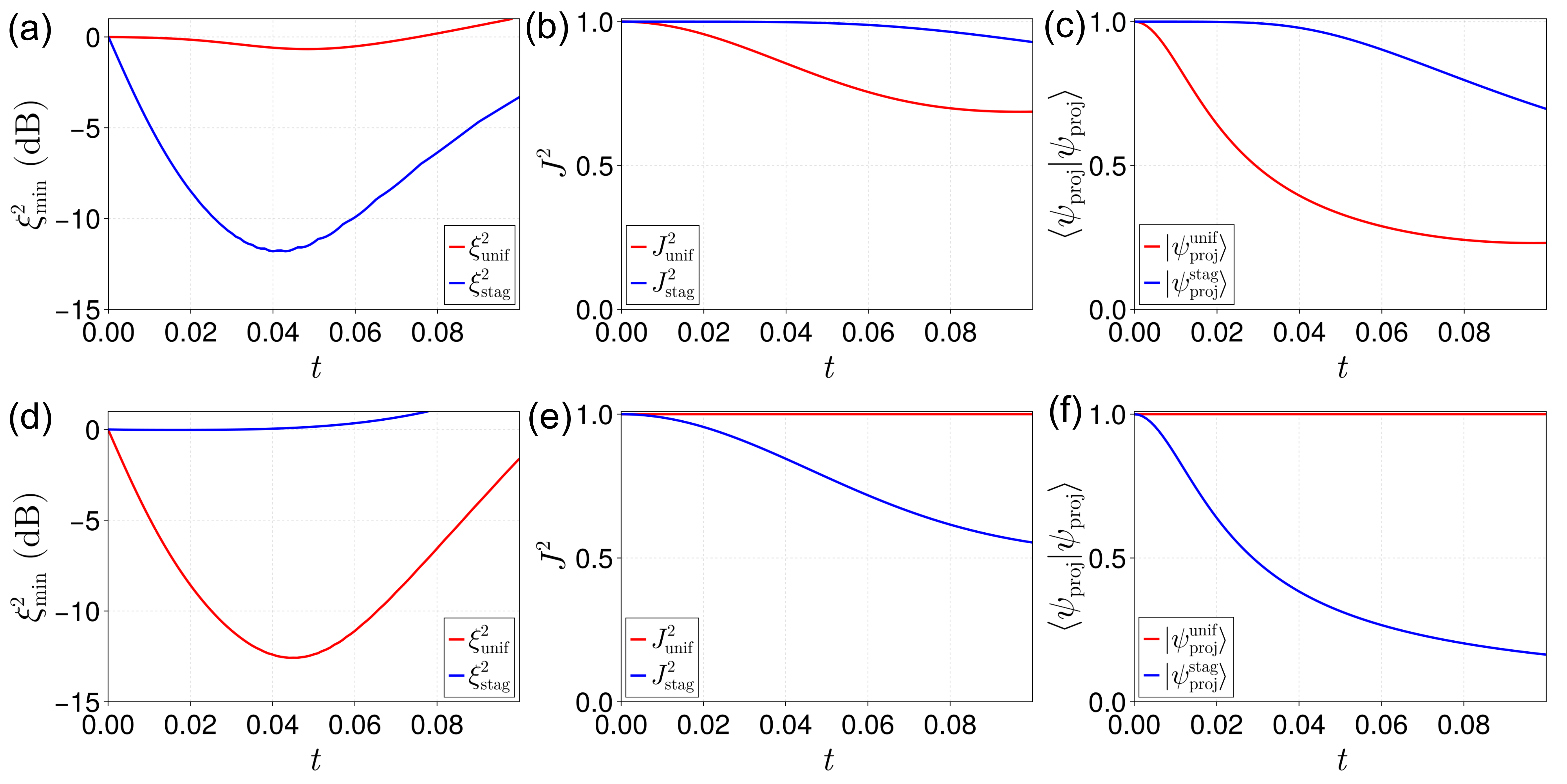}
\caption{Time evolution of (a) the eigenvalues of multiparameter squeezing matrix (in dB), (b) the mean value of $\mathbf{J}^{2}$ (divided by $(S_{1} + S_{2}) (S_{1} + S_{2} + 1)$ and (c) the projection norm $\langle \psi_{proj} \vert \psi_{proj} \rangle$ with $\hat{H}$ (red for uniform, blue for staggered) for $\chi_{12} = -2 \chi_{loc} $. (d,e,f) Same plots but with $\chi_{12} = 2 \chi_{loc}$. In all figures, $t$ is in $1/\chi_{\rm loc}$ unit.}
\label{figC}
\end{figure}

Hence, this is well defined for a spin of well-defined length $S$. However, when taking two spins of length $S_{1}$ and $S_{2}$, the total spin length of the total system can take any value between $\vert S_{1} - S_{2} \vert$ and $S_{1} + S_{2}$, and we need to make sure that we do not lose too much information by considering the Wigner distribution of a single spin length $S$. In our case, since we start with a product state of two CSS pointing in the same direction, we know that our initial state corresponds to a spin length $S = S_{1} + S_{2}$. However, the time evolved state might be a superposition of different spin length. Here, we will only consider the Wigner distribution corresponding to the larger Bloch sphere radius $\sqrt{S(S+1)}$ as it is the subspace in which we have the largest possible amount of squeezing. Therefore, in order to compute and plot the Wigner distribution of any time evolved state, first we need to project our state from the $\vert S_{1}, m_{1} \rangle \bigotimes \vert S_{2}, m_{2} \rangle$ basis to the basis of the collective spin of maximum length, $\vert S = S_{1} + S_{2}, m = S_{1} \pm S_{2} \rangle$ (depending if we look at the uniform or staggered spin). In order to construct the projector from one base to the other, we simply diagonalize numerically the following Hamiltonian:
\begin{equation}
    \hat{H}_{basis} = \lambda \mathbf{J}^{2}_{unif/stag} + \mu J^{z}_{unif/stag}
\end{equation}
written in the previous basis $\vert S_{1}, m_{1} \rangle \bigotimes \vert S_{2}, m_{2} \rangle$.

Finally, in order to check the validity of our approach, we compute the evolution in time of the norm of the projected state, as well as the mean value of $\mathbf{J}^{2}_{unif/stag}$, and show our results in Fig.~\ref{figC}. In the uniform case ($\chi_{12} = - 2 \chi_{loc}$), we see that the projection in the uniform subspace leaves the wavefunction unchanged, with $\mathbf{J}^{2}_{\rm unif} = S(S+1)$ ($S=S_{1} + S_{2}$) and $\langle \psi_{proj}^{\rm unif} \vert \psi_{proj}^{\rm unif} \rangle = 1$ at all times. On the other hand, in the staggered case ($\chi_{12} = 2 \chi_{loc}$), these quantities are not conserved, but at the time of optimal squeezing ($t \approx 0.04$), we still have $\mathbf{J}^{2}_{\rm unif}/(S(S+1) > 0.99$ and $\langle \psi_{proj}^{\rm stag} \vert \psi_{proj}^{\rm stag} \rangle > 0.97$.  These results show that at least up to the time of optimal squeezing, most of the weight of the wavefunction is indeed in the subspace of maximal spin length $S_{1} + S_{2}$, and confirm the validity of our approach to compute the Wigner distribution of the state.

\section{D- Classical trajectories}
\label{app_class_traj}

We give here some details on the derivation of the classical equation of motions given in Eq. \ref{class_traj}. In the semiclassical limit of very large spins, we can write the state of each spin in terms of the classical variables $z_{i}$ (mean value of $S^{z}_{i}$ and $\phi_{i}$ (azimuthal angle), such that   $\mathbf{S}_{i} = \begin{pmatrix}   \sqrt{1-z_{i}^{2}} \cos \left( \phi_{i} \right), & \sqrt{1-z_{i}^{2}} \sin \left( \phi_{i} \right), & z_{i}   \end{pmatrix}$. Assuming $z_{i}$ and $\phi_{i}$ satisfy the canonical relation $ \lbrace z_{i}, \phi_{i} \rbrace = 1$, we derive the classical Hamiltonian equations for the system:
\begin{align*}
    \frac{{\rm d} z_{1}}{{\rm d} t} &= N_{1} N_{2} \chi_{12} \sqrt{1 - (z_{1})^{2}} \sqrt{1 - (z_{2})^{2}} \sin \left( \phi_{1} - \phi_{2} \right), &\frac{{\rm d} \phi_{1}}{{\rm d} t} = - N_{1} N_{2} \chi_{12}  \sqrt{1 - (z_{2})^{2}} \frac{z_{1}}{\sqrt{1 - (z_{1})^{2}}} \cos \left( \phi_{1} - \phi_{2} \right) + 2 \chi_{loc} N_{1}^{2} z_{1} \\
    \frac{{\rm d} z_{2}}{{\rm d} t} &= N_{1} N_{2} \chi_{12} \sqrt{1 - (z_{1})^{2}} \sqrt{1 - (z_{2})^{2}} \sin \left( \phi_{2} - \phi_{1} \right)    &\frac{{\rm d} \phi_{2}}{{\rm d} t} = - N_{1} N_{2} \chi_{12}  \sqrt{1 - (z_{1})^{2}} \frac{z_{2}}{\sqrt{1 - (z_{2})^{2}}} \cos \left( \phi_{1} - \phi_{2} \right) + 2 \chi_{loc} N_{2}^{2} z_{2}
\end{align*}
We then introduce the collective variables $Z = z_{1} + z_{2}$, $\delta z = z_{1} - z_{2}$,  $\Phi = \phi_{1} + \phi_{2}$ and $\delta \phi = \phi_{1} - \phi_{2}$, and rewrite our equations of motion:
\begin{align*}
    \frac{{\rm d} Z}{{\rm d} t} &= 0 \\
     \frac{{\rm d} \delta z}{{\rm d} t} &= 2 N_{1} N_{2} \chi_{12} \sqrt{1 - \left(\frac{Z+\delta z}{2} \right)^{2}} \sqrt{1 - \left(\frac{Z-\delta z}{2} \right)^{2}} \sin \left( \delta \phi \right) \\
    \frac{{\rm d} \Phi}{{\rm d} t} &= - N_{1} N_{2} \chi_{12}  \left[ \sqrt{1 - \left(\frac{Z-\delta z}{2} \right)^{2}} \frac{Z+\delta z}{ \sqrt{4 -\left(Z+\delta z \right)^{2}}} + \sqrt{1 - \left(\frac{Z+\delta z}{2} \right)^{2}} \frac{Z-\delta z}{ \sqrt{4 -\left(Z-\delta z \right)^{2}}} \right] \cos \left( \delta \phi \right) + \chi_{loc} \left(  (N_{1}^{2} + N_{2}^{2}) Z + (N_{1}^{2} - N_{2}^{2}) \delta z \right) \\
    \frac{{\rm d} \delta \phi}{{\rm d} t} &= - N_{1} N_{2} \chi_{12}  \left[ \sqrt{1 - \left(\frac{Z-\delta z}{2} \right)^{2}} \frac{Z+\delta z}{ \sqrt{4 -\left(Z+\delta z \right)^{2}}} - \sqrt{1 - \left(\frac{Z+\delta z}{2} \right)^{2}} \frac{Z-\delta z}{ \sqrt{4 -\left(Z-\delta z \right)^{2}}} \right] \cos \left( \delta \phi \right) +  \chi_{loc}\left(  (N_{1}^{2} - N_{2}^{2}) Z + (N_{1}^{2} + N_{2}^{2}) \delta z \right).
\end{align*} 
With these new variables, it is clear that $Z$ is constant and that we can solve the whole dynamics by solving only the coupled equations for $\delta \phi$ and $\delta z$ since $\Phi$ does not appear in the derivatives. For our initial state, we have $z_{1} = z_{2} = 0$, with a quantum uncertainty proportional to $1/\sqrt{N_{1}}$ or $1/\sqrt{N_{2}}$, which vanishes in the large spin limit. Therefore, in the following, we will choose $Z_{0} = 0$, so that we have $Z=0$ at all times. Assuming further that $N_{1} = N_{2}$ we obtain for $\delta \phi$ and $\delta z$:
\begin{align*}
    \frac{{\rm d} \delta z}{{\rm d} t} &=  N_{1} N_{2} \chi_{12} \left(2 - \frac{\delta z^{2}}{2} \right) \sin \left( \delta \phi \right) \\
    \frac{{\rm d} \delta \phi}{{\rm d} t} &= - N_{1} N_{2} \chi_{12}  \delta z \cos \left( \delta \phi \right) + 2  N_{1} N_{2} \chi_{loc} \delta z.
\end{align*}
as well as ${\rm d} \Phi / {\rm d} t = 0$. With the same argument as before, applied to the $y$ component of the spin, we can set $\Phi_{0} = 0$ which completely constrains the dynamics of the system. Finally, under the assumption that $\chi_{12} = - 2 \chi_{loc}$, we recover the equations given in Eq. \ref{class_traj}, up to a rescaling of time $\tau = N_{1} N_{2} \chi_{12} t$.

\begin{figure}[ht!]
\includegraphics[width=0.48\textwidth]{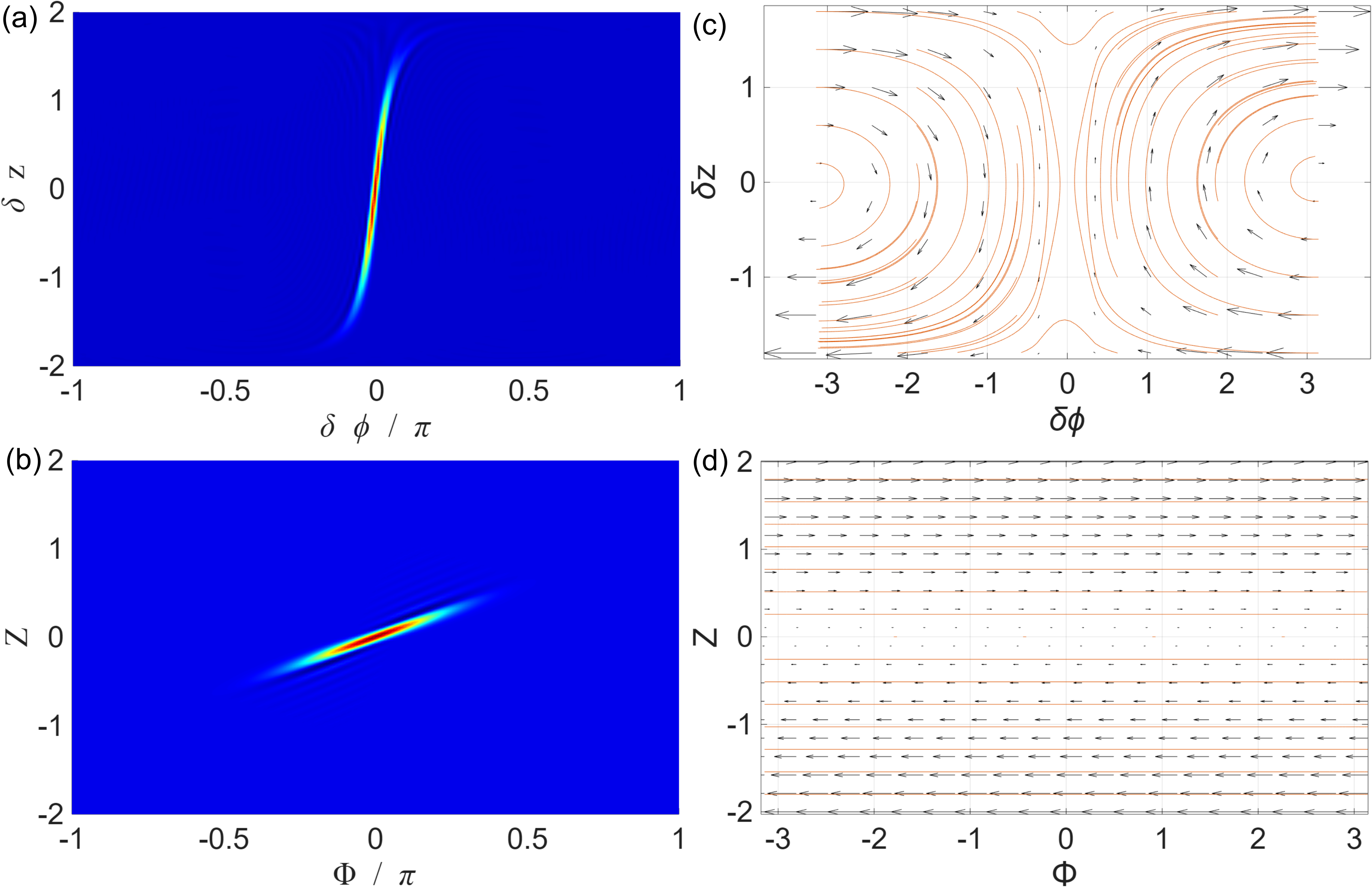}
\caption{ (a-b): Wigner distribution on the staggered Bloch sphere of the system at the time of the minimum of the squeezing with (a) $\chi_{12}= -2$ and (b) $\chi_{12} =2$. In both cases we have taken $N = 60$, $M=2$ and  $\chi_{loc} = 1$. (c-d): Semiclassical trajectories on the Bloch sphere associated with the previous values of $\chi_{12}$,  (c) $\chi_{12} = -2$ with the staggered collective spin and (d) $\chi_{12}=2$ with the uniform collective spin. Again we have taken $\chi_{loc} = 1$, $N=60$, $M=2$.
}
\label{figD}
\end{figure}

Using these equations, we can reconstruct the trajectories on the collective Bloch sphere for a given $\chi_{12}$, as shown in Fig.~\ref{figD}. In particular, in Fig.~\ref{figD} (c) we see that we have a continuous line of unstable fixed points at $\delta \phi = 0$, which is responsible for the antisqueezing of the spin along the $z$ axis. Moreover, we have a single stable fixed point at $\delta z =0, ~\delta \phi = \pi$.

\section*{E - Truncated cumulant expansion}

\subsection*{E1 - Definition and equations of motion}

In order to compute the time evolution of the system for a large number of tweezers ($M \geq 6$), we introduce an approximate numerical method, based on truncated cumulant expansion of the correlation function of the different spin observables. More specifically, we introduce the following basis of observables:
\begin{equation}
    T_{i}^{(m,m')} = \vert i, m \rangle \langle i, m' \vert
\end{equation}
where $\vert i, m \rangle$ denotes the eigenstate of $S^{z}_{i}$ of for the eigenvalue $m$, with fixed spin length $S = N/2$. These object obey the following commutation relations:
\begin{equation}
    \left[ T_{i}^{(m,m')}, T_{j}^{(n,n')}\right] = \delta_{ij} \left( \delta_{m' n} T_{i}^{(m,n')} - \delta_{m n'} T_{i}^{(n,m')}  \right).
\end{equation}
It is possible to express any spin observable in this basis as:
\begin{equation}
    S_{i}^{z} = \sum_{m=-S}^{S} m T_{i}^{(m,m)}, \quad S^{+}_{i} = \sum_{m=-S}^{S-1} \sqrt{S(S+1)-m(m+1)} T_{i}^{(m+1, m)}.
\end{equation}

From this, we can compute the time evolution of any mean value of these operators by computing its commutator with the Hamiltonian:
\begin{align}
    i \hbar \frac{{\rm d}}{{\rm d}t} \langle T_{i}^{(m,m')} \rangle &= \langle \left[ T_{i}^{(m,m')}, H\right] \rangle \nonumber \\
    &= \chi_{\rm loc} \left((m')^{2} - m^{2} \right) \langle T_{i}^{(m,m')} \rangle \nonumber \\
    &+ \frac{1}{2} \sum_{j\neq i} \chi_{ij} \sum_{l = -S+1}^{S} C^{-}(l) C^{-}(m') \langle T_{i}^{(m,m'-1)}T_{j}^{(l-1,l)}\rangle - C^{-}(l) C^{+}(m) \langle T_{i}^{(m+1,m')}T_{j}^{(l-1,l)} \rangle \nonumber \\
    &+ \frac{1}{2} \sum_{j\neq i} \chi_{ij} \sum_{l = -S}^{S-1} C^{+}(l) C^{+}(m') \langle T_{i}^{(m,m'+1)}T_{j}^{(l+1,l)} \rangle - C^{+}(l) C^{-}(m) \langle T_{i}^{(m-1,m')}T_{j}^{(l+1,l)}\rangle .
\end{align}
As one can see, the time evolution of these mean values involve higher order object, the two points correlations $\langle T_{i}^{(m,m')}T_{j}^{(n,n')}\rangle$. In turn, their evolution would involve third order terms and so on, until we reach order $M$. In order to keep a reasonable amount of terms in the temporal evolution, we here choose to truncate this expansion after the second order, i.e. we assume that all the cumulants of order 3 or higher are zero, leading to 
\begin{equation}
    \langle T_{i}^{(m,m'+1)}T_{j}^{(n,n')} T_{k}^{(p,p')} \rangle =  \langle T_{i}^{(m,m'+1)}T_{j}^{(n,n')} \rangle \langle T_{k}^{(p,p')} \rangle +  \langle T_{i}^{(m,m'+1)} T_{k}^{(p,p')}\rangle \langle T_{j}^{(n,n') }\rangle +  \langle T_{j}^{(n,n')}T_{k}^{(p,p')}\rangle \langle  T_{i}^{(m,m'+1)} \rangle - 2  \langle T_{i}^{(m,m'+1)} \rangle \langle T_{j}^{(n,n')} \rangle \langle T_{k}^{(p,p')} \rangle
\end{equation}
which effectively allows us to close our system of coupled equations after the second order. We call this 2nd order truncated cumulant expansion (TCE) \cite{Trifa24}.
This leads to a number of terms that scale as $O(M^{2} (N+1)^{4})$, which can even be reduced to $O((N+1)^{4})$ using the symmetries of our Hamiltonian. This approach is therefore very well suited to study a large number of subsystems $M$, but is limited to small $N$ within each subsystem.

\subsection*{E2 - Domain of validity and physical insight}

By construction, this method is numerically exact when we consider exactly $M=2$ subsystems. However, for $M=3$, we make an uncontrollable error when the third and higher order cumulants become non zero during the dynamics. Still, since the initial state is a uncorrelated state, this approximation is valid at $t=0$. We can therefore expect that our approximation remains satisfying at early time, with first 2nd order correlations (covariances) emerging, before higher order correlations can build up. In particular, squeezed states can be well approximated by gaussian states~\cite{PezzeRMP2018}, which are states where only the first two orders of cumulants are non zero. Therefore, this approximation should remain valid up to time comparable to the maximally squeezed state achieved in the dynamics, and we will verify this assumption in the next subsection.

Finally, we must stress out that unlike bosonic or fermionic system where all gaussian states can be mapped unto physical states, in the case of spin, there exist gaussian states that are highly non physical, for instance violating Heisenberg-Robertson inequality for spin. In order to avoid considering unphysical results, we compute at each time the ratio 
\begin{equation}
    \label{ratio}
    R = 4 {\rm Var} (J^{\rm max}) {\rm Var}  (J^{\rm min}) \rangle/\langle J^{x} \rangle^2
\end{equation}  
for the collective spin, with max, min corresponding here to the maximally and minimally squeezed direction in the $y, z$ plane, and consider only times until $R$ reaches a value below $1$, indicating a violation of the Heisenberg inequality, as we did in \cite{Trifa24}.

\subsection*{E3 - Benchmark versus exact diagonalization}

\begin{figure}[h!]
\includegraphics[width=0.9\textwidth]{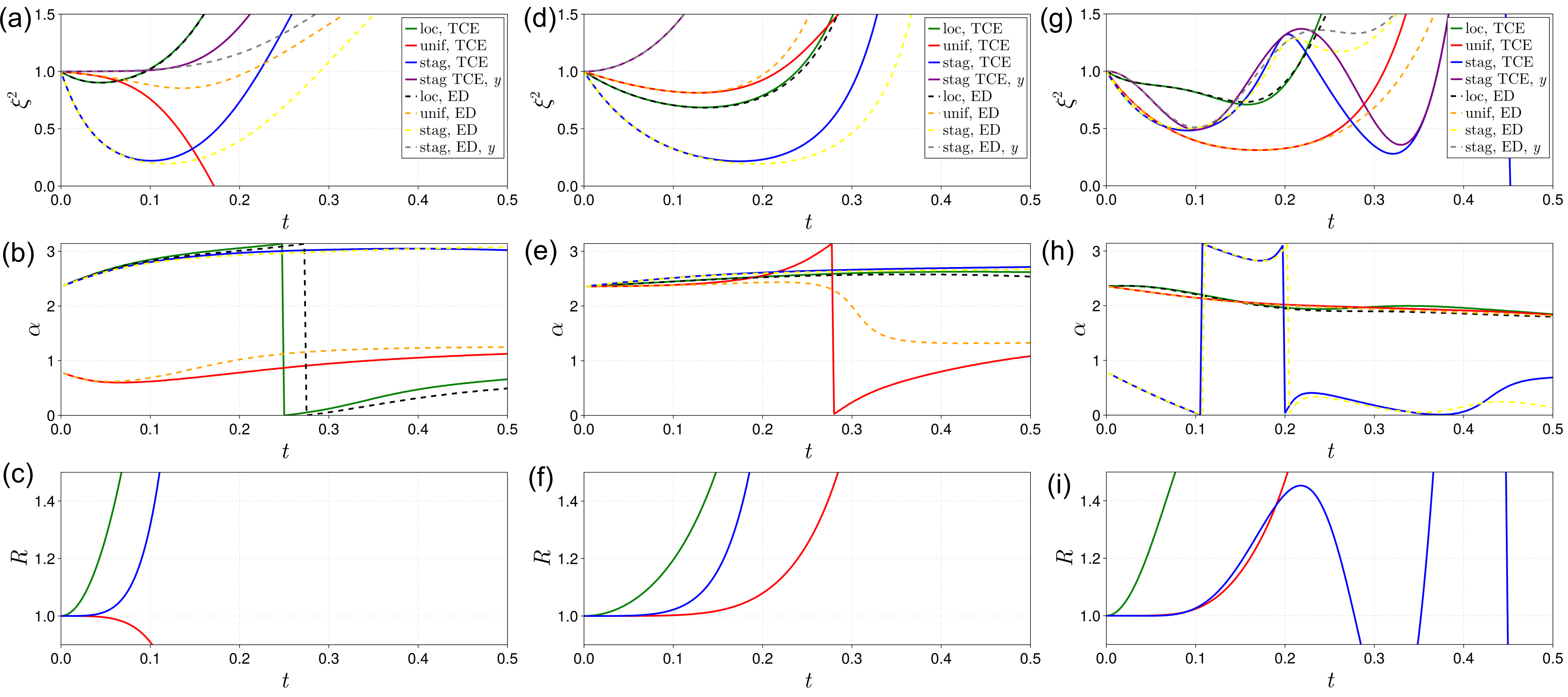}
\caption{ (a) Time evolution of the eigenvalues of multiparameter squeezing matrix (in dB), with $\hat{H}$ for $\chi_{12} = -2,  ~\chi_{loc} = 1 $, for $M=4$ and $N=6$. We show in green solid lines the local squeezing, in solid red the uniform one, in solid blue the staggered one and in dashed purple the staggered one at fixed angle (along $y$) for the TCE dynamics. We compare these results with the ones obtained with exact diagonalization (shown in dashed lines). (b) Time evolution of the angle of the optimal squeezing at each time. (c) Time evolution of $R$ (as defined in equation \ref{ratio}) for the local (green), uniform (red) and staggered (blue) spin operators. (e-i) Same figures as (a-c) with $\chi = -0.8$ (d-f) and $\chi = 2$ (g-i). In all figures we have taken $M=4$ and $N = 6$, and $t$ is in $1/\chi_{\rm loc}$ unit.}
\label{figE}
\end{figure}

We show here comparison between 2nd order TCE and ED for the time evolution of the cavity system with $M=4$ subsystems, with $N=10$ atoms in each. In particular, we show in Fig.~\ref{figE} that up to maximum time defined by $R$, we have a very good agreement between ED and TCE results, in particular until the time of optimal squeezing. This is true for various values of $\chi_{ij}$, with a good agreement for the quantity of interest (staggered squeezing), even if the agreement for the uniform squeezing can be worse for $< \chi_{ij} <$, but in this region of parameter we are not interested in the uniform squeezing at the is but small amount of squeezing for this configuration.

From the comparison between the TCE and ED results, we can clearly see that the parameter $R_{\rm unif / stag}$ is a good indication of the validity of the TCE approximation. Indeed, in case where $R$ remains strictly larger than $1$ and does present a local maximum, we have a very good quantitative agreement for associated the squeezing, at least until the time of the minimum of squeezing is reached (see fig. \ref{figE} (a) and (d) for the staggered squeezing, and (d) and (g) for the uniform one. On the contrary, as soon as $R$ reaches a local maximum (which preludes a rapid fall), we start observing strong disagreement between TCE and ED results (see Fig.~\ref{figE} (g)) for the staggered squeezing. Finally, for the uniform squeezing, it happens that $R$ takes from the start of the dynamics values below one, in which case we do not consider at all the results given by the TCE (see Fig.~\ref{figE} (a)). We also see a very good quantitative agreement for the squeezing angle up to the time of the minimum of the squeezing parameter, as long as the previous conditions on $R$ are satisfied ($R>1$ and before a local maximum is reached).

\section*{F - Time evolution of the squeezing parameter $M=20$}

\begin{figure}[h!]
\includegraphics[width=0.9\textwidth]{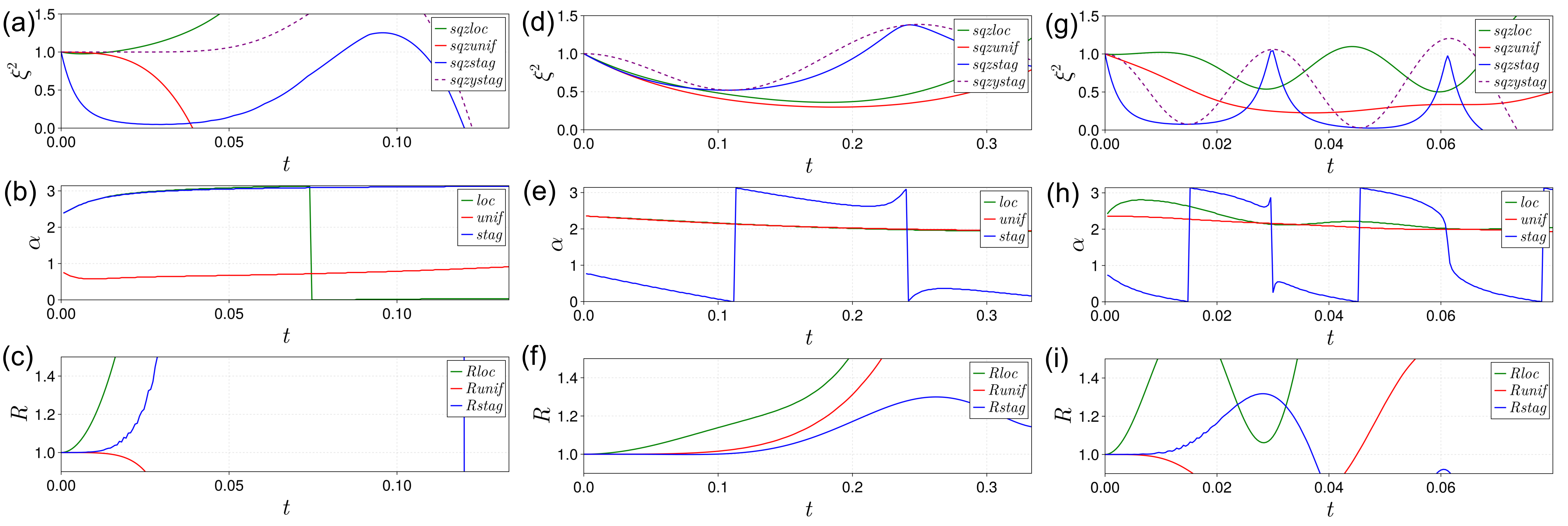}
\caption{ (a) Time evolution of the eigenvalues of multiparameter squeezing matrix (in dB), with $\hat{H}$ for $\chi_{12} = -2,  ~\chi_{loc} = 1 $. We show in green solid lines the local squeezing, in solid red the uniform one, in solid blue the staggered one and in dashed purple the staggered one at fixed angle (along $y$). (b) Time evolution of the angle of the optimal squeezing at each time. (c) Time evolution of $R$ (as defined in equation \ref{ratio}) for the local (green), uniform (red) and staggered (blue) spin operators. (e-i) Same figures as (a-c) with $\chi = 0.2$ (d-f) and $\chi = 4$ (g-i). In all figures we have taken $M=20$ and $N = 10$, and $t$ is in $1/\chi_{\rm loc}$ unit.}
\label{figF}
\end{figure}

In this section, we show and discuss more in details the results obtained for the time evolution of the squeezing parameter with the TCE simulations for $M=20$ ensembles. We show in Fig.~\ref{figF} the time evolution of the squeezing parameter, the optimal squeezing angle and the ratio $R$ (as defined in the previous section) for three distinct values of $\chi$ (with $\chi_{\rm loc} = 1$).

First, for $\chi = -2$, we have the highest amount of staggered squeezing, but no local squeezing and almost no staggered squeezing along the $y$ direction, even though the staggered squeezing angle is close to $\pi$. We can see that up to the minimum in time of the squeezing parameter, the ratio $R$ is well above 1. On the other hand, in the case of the uniform collective squeezing, it goes below one at short time, so we do not take into account the uniform squeezing in our results. From the comparison made with ED simulation, we know that for this specific value of $\chi$ we expect no uniform squeezing. Only for small values of $\chi$ (typically $-0.25<\chi<0.75$) we did find a clear minimum in the uniform squeezing parameter that occurs before a negative value of $R_{\rm unif}$. Moreover, only for small positive values of $\chi$ did we find an amount of uniform collective squeezing that was more important (for this value of $\chi$) than the amount of staggered collective squeezing (see Fig.~\ref{figF} (d) for $\chi = 0.2$), and therefore we choose not to take into account the uniform squeezing in Fig.~\ref{fig3}. For this specific value of $\chi$, the angle of uniform squeezing is comparable to the angle of OAT dynamics, as shown by the good agreement with the angle of local squeezing which obeys almost an exact OAT dynamics for this value of $\chi$.

Finally, for large positive $\chi$, the TCE dynamics is less stable for the staggered squeezing, as we obtain values of $R_{\rm stag}$ below one at shorter times, but the minimum of the squeezing is still reached before the $R_{\rm stag}$ reaches a local maximum, which is a prelude to a rapid fall of $R_{\rm stag}$, which reaches a value 1 shortly after this local maximum.

\section*{G - Reconstruction of the joint probability density}

In the case of a large number of ensembles $M=20$, it is not possible to reconstruct exactly the joint probability density ${\cal P} \left( m_{z}^{A}, m_{z}^{B}, t, \delta \theta, \epsilon \right)$. With the TCE simulation, we have access to the distribution probability of the spin projection within each tweezer ensemble $i$, $m_{z}^{(i)}$, as well as their variance and the covariances between these ensembles. Yet, reconstructing the projected distribution density on the collective spin projection in each subsystem $A$, $B$ from the local probabilities is a very difficult problem, as the number of possible outcome to compute is exponential with the number of ensembles $M$. Here, we choose to rely on a gaussian approximation of the collective probability density in each subsystem $A$, $B$. We first compute the mean value and variance of the spin operators $J^{z}_{A, B}$ as well as their covariance from the TCE simulation for a single realization of the common noise $\epsilon$. We then reconstruct the joint probability as a multivariate normal law:
\begin{equation}
    {\cal P}_{G} \left( m_{z}^{A}, m_{z}^{B}, t, \delta \theta, \epsilon \right) = \frac{1}{2 \pi} \frac{1}{\sqrt{{\rm det} \Sigma}} \exp \left[ - \frac{1}{2} (X-\mu)^{T} \Sigma^{-1} (X-\mu) \right]
\end{equation}
with $X = (m_{z}^{A}, \, m_{z}^{B})^{T}$, $\mu = ( \langle J^{z}_{A} \rangle, \, \langle J^{z}_{B} \rangle)^{T}$ and $\Sigma$ is the $2 \times 2$ covariance matrix, $\Sigma_{11} = {\rm Var} (J^{z}_{A})$, $\Sigma_{22} = {\rm Var} (J^{z}_{B})$ and $\Sigma_{12} = \Sigma_{21} = {\rm Cov} (J^{z}_{A}, \, J^{z}_{B})$. We then average this probability density over all the possible common noise realization to obtain the average probability, $\bar{\cal P}_{G}$:
\begin{equation}
    \bar{\cal P}_{G} =  \frac{1}{2 \pi} \int_{0}^{2 \pi} \mathrm{d} \epsilon \, {\cal P}_{G} \left( m_{z}^{A}, m_{z}^{B}, t, \delta \theta, \epsilon \right).
\end{equation}
Finally, we did check that for small number of ensembles, this approximation was valid at short times (it is exact for $t=0$, i.e. for coherent spin states), and remains valid as long as the state is not too squeezed, which is the case for the optimal state for differential measurement, which occurs well before the maximally squeezed state.
 

\begin{thebibliography}{99}

\bibitem{CroninRMP2009}
A. D. Cronin, J. Schmiedmayer, and D. E. Pritchard,
Optics and interferometry with atoms and molecules,
\textit{Rev. Mod. Phys.} \textbf{81}, 1051 (2009).

\bibitem{Poli2013}
N. Poli, C. W. Oates, P. Gill, and G. M. Tino,
Optical atomic clocks,
\textit{Riv. Nuovo Cimento} \textbf{36}, 555 (2013).

\bibitem{LudlowRMP2015}
A. D. Ludlow, M. M. Boyd, J. Ye, E. Peik, and P. O. Schmidt,
Optical atomic clocks,
\textit{Rev. Mod. Phys.} \textbf{87}, 637 (2015).

\bibitem{BongsNatRev2019}
K. Bongs, M. Holynski, J. Vovrosh, et al.,
Taking atom interferometric quantum sensors from the laboratory to real-world applications,
\textit{Nat. Rev. Phys.} \textbf{1}, 731 (2019).

\bibitem{RosiNature2014}
G. Rosi, F. Sorrentino, L. Cacciapuoti, et al.,
Precision measurement of the Newtonian gravitational constant using cold atoms,
\textit{Nature} \textbf{510}, 518 (2014).

\bibitem{Janvier2022}
C. Janvier, V. M\'enoret, B. Desruelle, S. Merlet, A. Landragin, and F. Pereira dos Santos,
Compact differential gravimeter at the quantum projection-noise limit,
\textit{Phys. Rev. A} \textbf{105}, 022801 (2022).

\bibitem{PezzeRMP2018}
L. Pezz\`e, A. Smerzi, M. K. Oberthaler, R. Schmied, and P. Treutlein,
Quantum metrology with nonclassical states of atomic ensembles,
\textit{Rev. Mod. Phys.} \textbf{90}, 035005 (2018).

\bibitem{Salvi2018}
L. Salvi, N. Poli, V. Vuleti\'c, and G. M. Tino,
Squeezing on momentum states for atom interferometry,
\textit{Phys. Rev. Lett.} \textbf{120}, 033601 (2018).

\bibitem{Szigeti2020}
S. S. Szigeti, S. P. Nolan, J. D. Close, and S. A. Haine,
High-precision quantum-enhanced gravimetry with a Bose--Einstein condensate,
\textit{Phys. Rev. Lett.} \textbf{125}, 100402 (2020).

\bibitem{CorgierPRL2021}
R. Corgier, N. Gaaloul, A. Smerzi, and L. Pezz\`e,
Delta-kick squeezing,
\textit{Phys. Rev. Lett.} \textbf{127}, 183401 (2021).

\bibitem{Colombo2022}
S. Colombo, E. Pedrozo-Pe\~nafiel, and V. Vuleti\'c,
Entanglement-enhanced optical atomic clocks,
\textit{Appl. Phys. Lett.} \textbf{121}, 210502 (2022).

\bibitem{RobinsonNatPhys2024}
J. M. Robinson, M. Miklos, Y. M. Tso, et al.,
Direct comparison of two spin-squeezed optical clock ensembles at the $10^{-17}$ level,
\textit{Nat. Phys.} \textbf{20}, 208 (2024).

\bibitem{Yang2025}
Y. A. Yang, M. Miklos, Y. M. Tso, S. Kraus, J. Hur, and J. Ye,
Clock precision beyond the standard quantum limit at the $10^{-18}$ level,
\textit{Phys. Rev. Lett.} \textbf{135}, 193202 (2025).

\bibitem{OckeloenPRL2013}
C. F. Ockeloen, R. Schmied, M. F. Riedel, and P. Treutlein,
Quantum metrology with a scanning probe atom interferometer,
\textit{Phys. Rev. Lett.} \textbf{111}, 143001 (2013).

\bibitem{Muessel2014}
W. Muessel, H. Strobel, D. Linnemann, D. B. Hume, and M. K. Oberthaler,
Scalable spin squeezing for quantum-enhanced magnetometry with Bose--Einstein condensates,
\textit{Phys. Rev. Lett.} \textbf{113}, 103004 (2014).

\bibitem{Greve2022}
G. P. Greve, C. Luo, B. Wu, et al.,
Entanglement-enhanced matter-wave interferometry in a high-finesse cavity,
\textit{Nature} \textbf{610}, 472 (2022).

\bibitem{Cassens2025}
C. Cassens, B. Meyer-Hoppe, E. Rasel, and C. Klempt,
Entanglement-enhanced atomic gravimeter,
\textit{Phys. Rev. X} \textbf{15}, 011029 (2025).

\bibitem{KomarNATPHYS2014}
P. K\'om\'ar, E. M. Kessler, M. Bishof, L. Jiang, A. S. S\o rensen, J. Ye, and M. D. Lukin,
A quantum network of clocks,
\textit{Nat. Phys.} \textbf{10}, 582 (2014).

\bibitem{malia_nat_2022}
B. K. Malia, Y. Wu, J. Mart\'inez-Rinc\'on, and M. A. Kasevich,
Distributed quantum sensing with mode-entangled spin-squeezed atomic states,
\textit{Nature} \textbf{612}, 661 (2022).

\bibitem{Nichol2022}
B. C. Nichol, R. Srinivas, D. Nadlinger, P. Drmota, D. Main, G. Araneda, C. J. Ballance, and D. M. Lucas,
An elementary quantum network of entangled optical atomic clocks,
\textit{Nature} \textbf{609}, 689 (2022).



\bibitem{LiScience2026}
Y. Li, L. Joosten, Y. Baamara, P. Colciaghi, A. Sinatra, P. Treutlein, and T. Zibold,
Multiparameter estimation with an array of entangled atomic sensors,
\textit{Science} \textbf{391}, 374 (2026).

\bibitem{Dietze2026}
K. Dietze, L. Pelzer, L. Krinner, F. Dawel, J. Kramer, N. C. H. Spethmann, T. Kielinski, K. Hammerer, K. Stahl, J. Klose, S. D\"orscher, C. Lisdat, E. Benkler, and P. O. Schmidt,
Entanglement-enhanced optical ion clock,
\textit{Phys. Rev. Lett.} \textbf{136}, 073601 (2026).

\bibitem{YeZollerPRL2024}
J. Ye and P. Zoller,
Essay: Quantum sensing with atomic, molecular, and optical platforms for fundamental physics,
\textit{Phys. Rev. Lett.} \textbf{132}, 190001 (2024).

\bibitem{PezzeArXiv2025}
L. Pezz\`e and A. Smerzi,
Advances in multiparameter quantum sensing and metrology,
\textit{arXiv:2502.17396} (2025).

\bibitem{Apellaniz2018}
I. Apellaniz, I. Urizar-Lanz, Z. Zimbor\'as, P. Hyllus, and G. T\'oth,
Precision bounds for gradient magnetometry with atomic ensembles,
\textit{Phys. Rev. A} \textbf{97}, 053603 (2018).

\bibitem{Stray2022}
B. Stray, A. Lamb, A. Kaushik, J. Vovrosh, A. Rodgers, J. Winch, F. Hayati, D. Boddice, A. Stabrawa, A. Niggebaum, M. Langlois, Y.-H. Lien, S. Lellouch, S. Roshanmanesh, K. Ridley, G. de Villiers, G. Brown, T. Cross, G. Tuckwell, A. Faramarzi, N. Metje, K. Bongs, and M. Holynski,
Quantum sensing for gravity cartography,
\textit{Nature} \textbf{602}, 590 (2022).

\bibitem{Dickerson2013}
S. M. Dickerson, J. M. Hogan, A. Sugarbaker, D. M. S. Johnson, and M. A. Kasevich,
Multiaxis inertial sensing with long-time point source atom interferometry,
\textit{Phys. Rev. Lett.} \textbf{111}, 083001 (2013).

\bibitem{Templier2022}
S. Templier, P. Cheiney, Q. d'Armagnac de Castanet, B. Gouraud, H. Porte, F. Napolitano, P. Bouyer, B. Battelier, and B. Barrett,
Tracking the vector acceleration with a hybrid quantum accelerometer triad,
\textit{Sci. Adv.} \textbf{8}, eadd3854 (2022).

\bibitem{Stolzenberg2025}
K. Stolzenberg, C. Struckmann, S. Bode, R. Li, A. Herbst, V. Vollenkemper, D. Thomas, A. Rajagopalan, E. M. Rasel, N. Gaaloul, and D. Schlippert,
Multi-axis inertial sensing with two-dimensional matter-wave arrays,
\textit{Phys. Rev. Lett.} \textbf{134}, 143601 (2025).

\bibitem{EcknerNATURE2023}
W. J. Eckner, N. Darkwah Oppong, A. Cao, A. W. Young, W. R. Milner,
J. M. Robinson, J. Ye, and A. M. Kaufman,
Realizing spin squeezing with Rydberg interactions in an optical clock,
\textit{Nature} \textbf{621}, 734 (2023).

\bibitem{Zheng2022}
X. Zheng, J. Dolde, V. Lochab, B. N. Merriman, H. Li, and S. Kolkowitz,
Differential clock comparisons with a multiplexed optical lattice clock,
\textit{Nature} \textbf{602}, 425 (2022).

\bibitem{BaynhamNature2026}
C. F. A. Baynham, R. Hobson, O. Buchm\"uller, et al.,
A prototype differential atom interferometer for fundamental physics,
\textit{Nature} \textbf{654}, 622 (2026).

\bibitem{KitsonPRR2026}
P. Kitson, W. J. Chetcuti, G. Birkl, L. Amico, and J. Polo,
Rydberg atoms for electric-field gradiometry,
\textit{Phys. Rev. Research} \textbf{8}, 033010 (2026).

\bibitem{gessner_multiparameter_2020}
M. Gessner, A. Smerzi, and L. Pezz\`e,
Multiparameter squeezing for optimal quantum enhancements in sensor networks,
\textit{Nat. Commun.} \textbf{11}, 3817 (2020).

\bibitem{Pezz2025}
L. Pezz\`e and A. Smerzi,
Distributed quantum multiparameter estimation with optimal local measurements,
\textit{Phys. Rev. Lett.} \textbf{135}, 260805 (2025).

\bibitem{Bate2025}
J. Bate, A. Hamann, M. Canteri, A. Winkler, Z. X. Koong, V. Krutyanskiy, W. D\"ur, and B. P. Lanyon,
Experimental distributed quantum sensing in a noisy environment,
\textit{Phys. Rev. Lett.} \textbf{135}, 220801 (2025).


\bibitem{Picot2026}
T. Picot, C. Raphin, M. Kern, P.-A. Bourdel, T. Ahamdach, J. Reichel, and R. Long,
Extended single-atom tweezer arrays in high-cooperativity cavity-QED,
\textit{arXiv:2607.21515} (2026).

\bibitem{CooperPRX2018}
A. Cooper, J. P. Covey, I. S. Madjarov, S. G. Porsev, M. S. Safronova, and M. Endres,
Alkaline-earth atoms in optical tweezers,
\textit{Phys. Rev. X} \textbf{8}, 041055 (2018).

\bibitem{YoungNat2020}
A. W. Young, W. J. Eckner, W. R. Milner, et al.,
Half-minute-scale atomic coherence and high relative stability in a tweezer clock,
\textit{Nature} \textbf{588}, 408 (2020).

\bibitem{KaufmanNatPhys2021}
A. M. Kaufman and K.-K. Ni,
Quantum science with optical tweezer arrays of ultracold atoms and molecules,
\textit{Nat. Phys.} \textbf{17}, 1324 (2021).

\bibitem{ManteschNat2025}
H. J. Manetsch, G. Nomura, E. Bataille, et al.,
A tweezer array with 6,100 highly coherent atomic qubits,
\textit{Nature} \textbf{647}, 60 (2025).

\bibitem{Periwal2021}
A. Periwal, E. S. Cooper, P. Kunkel, J. F. Wienand, E. J. Davis, and M. Schleier-Smith,
Programmable interactions and emergent geometry in an array of atom clouds,
\textit{Nature} \textbf{600}, 630 (2021).

\bibitem{GlicensteinPRA2021}
A. Glicenstein, G. Ferioli, L. Brossard, Y. R. P. Sortais, D. Barredo, F. Nogrette, I. Ferrier-Barbut, and A. Browaeys,
Preparation of one-dimensional chains and dense cold atomic clouds with a high-numerical-aperture four-lens system,
\textit{Phys. Rev. A} \textbf{103}, 043301 (2021).

\bibitem{Cooper2024}
E. S. Cooper, P. Kunkel, A. Periwal, and M. Schleier-Smith,
Graph states of atomic ensembles engineered by photon-mediated entanglement,
\textit{Nat. Phys.} \textbf{20}, 770 (2024).

\bibitem{SeubertPRX2025}
M. Seubert, L. Hartung, S. Welte, G. Rempe, and E. Distante,
Tweezer-assisted subwavelength positioning of atomic arrays in an optical cavity,
\textit{PRX Quantum} \textbf{6}, 010322 (2025).

\bibitem{Davis2016}
E. Davis, G. Bentsen, and M. Schleier-Smith,
Approaching the Heisenberg limit without single-particle detection,
\textit{Phys. Rev. Lett.} \textbf{116}, 053601 (2016).

\bibitem{Frowis2016}
F. Fr\"owis, P. Sekatski, and W. D\"ur,
Detecting large quantum Fisher information with finite measurement precision,
\textit{Phys. Rev. Lett.} \textbf{116}, 090801 (2016).

\bibitem{Macri2016}
T. Macr\`i, A. Smerzi, and L. Pezz\`e,
Loschmidt echo for quantum metrology,
\textit{Phys. Rev. A} \textbf{94}, 010102 (2016).

\bibitem{Hosten2016Nat}
O. Hosten, N. Engelsen, R. Krishnakumar, et al.,
Measurement noise 100 times lower than the quantum-projection limit using entangled atoms,
\textit{Nature} \textbf{529}, 505 (2016).

\bibitem{Kaubruegger2026}
R. Kaubruegger, D. Fallas Padilla, A. Shankar, C. Hotter, S. R. Muleady, J. Bringewatt, Y. Baamara, E. Abbasgholinejad, A. V. Gorshkov, K. M\o lmer, J. K. Thompson, and A. M. Rey,
Lieb--Mattis states for robust entangled differential phase sensing,
\textit{Phys. Rev. X} \textbf{16}, 021052 (2026).

\bibitem{Chu2026}
A. Chu, M. Mamaev, M. Koppenh\"ofer, M. Yuan, and A. A. Clerk,
Reconfigurable dissipative entanglement between many spin ensembles: From robust quantum sensing to many-body state engineering,
\textit{Phys. Rev. X} \textbf{16}, 021047 (2026).

\bibitem{LerouxPRL2010}
I. D. Leroux, M. H. Schleier-Smith, and V. Vuleti\'c,
Implementation of cavity squeezing of a collective atomic spin,
\textit{Phys. Rev. Lett.} \textbf{104}, 073602 (2010).

\bibitem{HostenScience2016b}
O. Hosten, R. Krishnakumar, N. J. Engelsen, and M. A. Kasevich,
Quantum phase magnification,
\textit{Science} \textbf{352}, 1552 (2016).

\bibitem{PedrozoNature2020}
E. Pedrozo-Pe\~nafiel, S. Colombo, C. Shu, et al.,
Entanglement on an optical atomic-clock transition,
\textit{Nature} \textbf{588}, 414 (2020).

\bibitem{kitagawa_squeezed_1993}
M. Kitagawa and M. Ueda,
Squeezed spin states,
\textit{Phys. Rev. A} \textbf{47}, 5138 (1993).

\bibitem{Corgier2023}
R. Corgier, M. Malitesta, A. Smerzi, and L. Pezz\`e,
Quantum-enhanced differential atom interferometers and clocks with spin-squeezing swapping,
\textit{Quantum} \textbf{7}, 965 (2023).

\bibitem{Malitesta2023}
M. Malitesta, A. Smerzi, and L. Pezz\`e,
Distributed quantum sensing with squeezed-vacuum light in a configurable array of Mach--Zehnder interferometers,
\textit{Phys. Rev. A} \textbf{108}, 032621 (2023).

\bibitem{Mamaev2025}
M. Mamaev, M. Koppenh\"ofer, A. Pocklington, and A. A. Clerk,
Non-Gaussian generalized two-mode squeezing: Applications to two-ensemble spin squeezing and beyond,
\textit{Phys. Rev. Lett.} \textbf{134}, 073603 (2025).

\bibitem{winelandPRA1994}
D. J. Wineland, J. J. Bollinger, W. M. Itano, and D. J. Heinzen,
Squeezed atomic states and projection noise in spectroscopy,
\textit{Phys. Rev. A} \textbf{50}, 67 (1994).

\bibitem{KajtochPRA2015}
D. Kajtoch and E. Witkowska,
Quantum dynamics generated by the two-axis countertwisting Hamiltonian,
\textit{Phys. Rev. A} \textbf{92}, 013623 (2015).

\bibitem{Luo2025}
C. Luo, H. Zhang, A. Chu, C. Maruko, A. M. Rey, and J. K. Thompson,
Hamiltonian engineering of collective XYZ spin models in an optical cavity,
\textit{Nat. Phys.} \textbf{21}, 916 (2025).

\bibitem{Trifa24}
Y. Trifa and T. Roscilde,
Scalable spin squeezing in two-dimensional arrays of dipolar large-$S$ spins,
\textit{Phys. Rev. Lett.} \textbf{133}, 083601 (2024).

\bibitem{GanderBIT1994}
W. Gander, G. H. Golub, and R. Strebel,
Least-squares fitting of circles and ellipses,
\textit{BIT Numer. Math.} \textbf{34}, 558 (1994).

\bibitem{FosterOL02}
G. T. Foster, J. B. Fixler, J. M. McGuirk, and M. A. Kasevich,
Method of phase extraction between coupled atom interferometers using ellipse-specific fitting,
\textit{Opt. Lett.} \textbf{27}, 951 (2002).

\bibitem{CorgierQST2025}
R. Corgier, M. Malitesta, L. A. Sidorenkov, F. Pereira dos Santos, G. Rosi, G. M. Tino, A. Smerzi, L. Salvi, and L. Pezz\`e,
Optimized squeezing for accurate differential sensing under large phase noise,
\textit{Quantum Sci. Technol.} \textbf{10}, 045016 (2025).

\bibitem{varilly_moyal_1989}
J. C. V\'arilly and J. M. Gracia-Bond\'ia,
The Moyal representation for spin,
\textit{Ann. Phys.} \textbf{190}, 107 (1989).

\bibitem{klimov_generalized_2017}
A. B. Klimov, J. L. Romero, and H. de Guise,
Generalized $SU(2)$ covariant Wigner functions and some of their applications,
\textit{J. Phys. A: Math. Theor.} \textbf{50}, 323001 (2017).

\end{thebibliography}
\end{document}